\documentclass[aps,pra,reprint,superscriptaddress]{revtex4-1}
\usepackage[utf8]{inputenc}

\usepackage{graphicx}  
\usepackage{xcolor}
\usepackage{bm}        

\usepackage{amssymb} 
\usepackage{amsmath}
\usepackage{amsthm}
\usepackage{amsfonts}
\usepackage{mathtools}
\usepackage{ragged2e}
\usepackage{tikz}
\usepackage{hyperref}

\usepackage[capitalize]{cleveref}
\crefname{section}{Sec.}{Secs.}

\usepackage{bbm}
\usepackage{todonotes}

\theoremstyle{plain}  

\newtheorem{thm}{Theorem}[section]
\newtheorem{lem}[thm]{Lemma}

\theoremstyle{definition}

\newcommand{\ket}[1]{\left|#1\right\rangle}
\newcommand{\bra}[1]{\left\langle#1\right|}
\newcommand{\He}{H \mkern-2mu e}
\newcommand{\ssdirty}{\textsc{SelSwapDirty}~}

\usepackage{color}
\usepackage{xcolor}
\definecolor{darkred}{RGB}{150,0,0}
\definecolor{darkgreen}{RGB}{0,150,0}
\definecolor{darkblue}{RGB}{0,0,150}
\definecolor{grey}{RGB}{150,150,150}

\newcommand{\mcM}{\mathcal{M}}
\newcommand{\mcN}{\mathcal{N}}

\newcommand{\mcV}{\mathcal{V}}
\newcommand{\mcW}{\mathcal{W}}

\newcommand{\mbbF}{\mathbb{F}}

\newcommand{\mbbN}{\mathbb{N}}

\begin{document}
	\title{Initial state preparation for quantum chemistry on quantum computers}
	
	\author{Stepan Fomichev}
	\thanks{These two authors contributed equally.\\
		stepan.fomichev@xanadu.ai\\
		kasra.hejazi@xanadu.ai}
	\affiliation{Xanadu, Toronto, ON, M5G2C8, Canada}
	\author{Kasra Hejazi}
	\thanks{These two authors contributed equally.\\
		stepan.fomichev@xanadu.ai\\
		kasra.hejazi@xanadu.ai}
	\affiliation{Xanadu, Toronto, ON, M5G2C8, Canada}
	\author{Modjtaba Shokrian Zini}
	\affiliation{Xanadu, Toronto, ON, M5G2C8, Canada}
	\author{Matthew Kiser}
	\affiliation{Volkswagen AG, Berliner Ring 2, 38440 Wolfsburg, Germany}
	\affiliation{TUM School of Natural Sciences, Technical University of
		Munich, Garching, Germany}
	\author{Joana Fraxanet}
	\affiliation{Xanadu, Toronto, ON, M5G2C8, Canada}
	\affiliation{
		ICFO - Institut de Ci\`encies Fot\`oniques, The Barcelona Institute of Science and Technology, 08860 Castelldefels (Barcelona), Spain}
	\author{Pablo Antonio Moreno Casares}
	\affiliation{Xanadu, Toronto, ON, M5G2C8, Canada}
	\author{Alain Delgado}
	\affiliation{Xanadu, Toronto, ON, M5G2C8, Canada}
	\author{Joonsuk Huh}
	\affiliation{Xanadu, Toronto, ON, M5G2C8, Canada}
	\affiliation{Department of Chemistry, Sungkyunkwan University, Suwon 16419, Republic of Korea}
	\affiliation{SKKU Advanced Institute of Nanotechnology (SAINT), Sungkyunkwan University, Suwon 16419, Korea}
	\affiliation{Institute of Quantum Biophysics, Sungkyunkwan University, Suwon 16419, Korea}
	\author{Arne-Christian Voigt}
	\affiliation{Volkswagen AG, Berliner Ring 2, 38440 Wolfsburg, Germany}
	\author{Jonathan E. Mueller}
	\affiliation{Volkswagen AG, Berliner Ring 2, 38440 Wolfsburg, Germany}
	\author{Juan Miguel Arrazola}
	\affiliation{Xanadu, Toronto, ON, M5G2C8, Canada}

	\begin{abstract}
		Quantum algorithms for ground-state energy estimation of chemical systems require a high-quality initial state. 
		However, initial state preparation is commonly either neglected entirely, or assumed to be solved by a simple product state like Hartree-Fock.  
		Even if a nontrivial state is prepared, strong correlations render ground state overlap inadequate for quality assessment. 
		In this work, we address the initial state preparation problem with an end-to-end algorithm that \textit{prepares} and \textit{quantifies} the quality of initial states, accomplishing the latter with a new metric -- the energy distribution.
		To be able to prepare more complicated initial states, we introduce an implementation technique for states in the form of a sum of Slater determinants that exhibits significantly better scaling than all prior approaches.
		We also propose low-precision quantum phase estimation (QPE) for further state quality refinement. 
		The complete algorithm is capable of generating high-quality states for energy estimation, and is shown in select cases to lower the overall estimation cost by several orders of magnitude when compared with the best single product state ansatz. 
		More broadly, the energy distribution picture suggests that the goal of QPE should be reinterpreted as generating improvements compared to the energy of the initial state and other classical estimates, which can still be achieved even if QPE does not project directly onto the ground state. Finally, we show how the energy distribution can help in identifying potential quantum advantage. 
	\end{abstract}
	
	\maketitle

	\section{Introduction}
	
	One of the main contenders for useful applications of quantum computers is the simulation of many-body physics, in particular for quantum chemistry and materials science. Of special interest is the determination of ground-state energies, which have broad application \cite{aspuru2005simulated, reiher2017elucidating, cao2019quantum, von2021quantum, ho2018promise,rice2021quantum,clinton2022towards,delgado2022simulating,bauer2020quantum}. Many different quantum methods for ground-state energy determination have been proposed, ranging from quantum phase estimation (QPE) and its variants \cite{kitaev1995quantum,nielsen2010quantum,cleve1998quantum,knill2007optimal,poulin2009sampling}, to more recent developments \cite{ge2019faster,lin2022heisenberg,lin2020near}. We refer to these methods as \textit{quantum energy estimation} algorithms. Each of these methods requires a high quality initial state, where quality is traditionally understood in terms of the overlap with the ground state. The quality of the initial state directly impacts the performance and runtime of any energy estimation algorithm, making it crucial to develop advanced methods for initial state preparation. 
	
	While decisive for the success of quantum algorithms, initial state preparation is often treated as separate from the energy estimation algorithm, and has not received as much attention as other aspects of quantum algorithms in the literature. 
	A common approach to preparing an initial state is to take an approximate wavefunction from a traditional quantum chemistry method and encode it on a quantum computer. 
	The Hartree-Fock state is the simplest and computationally cheapest choice. 
	Even though it has been found to have high overlap with the ground state in small molecules \cite{reiher2017elucidating, tubman2018postponing, veis2012relativistic, von2021quantum}, it is seriously lacking for strongly-correlated systems \cite{kohn1999nobel}, such as molecules with stretched bonds \cite{wang2008quantum, veis2010quantum, veis2012relativistic, veis2014adiabatic, sugisaki2016quantum, tubman2018postponing} and more complex molecules with transition-metal centers~\cite{lee2023evaluating}. 
	Beyond the Hartree-Fock state, a variety of methods have been proposed to encode sums of Slater determinants (SOS) \cite{ortiz2001quantum, wang2009efficient, babbush2015chemical, tubman2018postponing, bravyi2019approximation} or matrix-product states (MPS) \cite{schon2005sequential,malz2023preparation}. Ground-state energy estimation was explored using SOS states obtained from configuration interaction singles and doubles \cite{babbush2015chemical}, active space methods \cite{wang2008quantum, veis2010quantum, veis2012relativistic}, and selective configuration interaction methods \cite{tubman2018postponing}. However, the performance of these approaches has mostly not been evaluated or compared beyond small, uncorrelated molecules. 
	
	Other approaches that can be categorized as quantum heuristic methods have also been considered widely. While adiabatic state preparation is likely the most well-known \cite{farhi2001quantum, farhi2000numerical, wu2002polynomial, aspuru2005simulated, albash2018adiabatic}, other heuristics include variational methods \cite{yung2014transistor, peruzzo2014variational, romero2018strategies, dallaire2019low} and quantum imaginary time evolution \cite{motta2020determining, gomes2020efficient,kamakari2022digital, hejazi2023adiabatic}.  While promising, to date most of these methods have been demonstrated only for small molecules \cite{aspuru2005simulated, veis2014adiabatic, peruzzo2014variational, romero2018strategies, dallaire2019low}, and suffer from various shortcomings such as long runtimes, expensive classical optimization, or costly state tomography. More broadly, the absence of any guarantee of their success in state preparation is problematic.
	
	The variety of state preparation approaches raises the question of which method is best suited to which situations. Furthermore, it is not even clear how one should compare different possible state preparation schemes for actual problems of interest. For example, in general the overlap with the ground state cannot serve as a practical metric for comparison or quality assessment since we typically do not know what the ground state is. These issues also make it difficult to quantify the total runtime of quantum algorithms and to understand their actual potential to outperform classical methods. Overall, there is a need for a framework that encompasses the most powerful methods for initial state preparation, provides tools to evaluate their quality, and allows us to make informed statements about the prospects for quantum advantage.

	In this work, we present a complete algorithm for preparing high-quality states for quantum energy estimation. Our state preparation algorithm begins with using quantum chemistry methods to obtain classical descriptions of approximate ground-state wavefunctions, either in SOS or MPS form. We then introduce a novel quantum algorithm for preparing SOS states on a quantum computer, with a better cost compared to all previous methods. This is complemented with resource estimation formulas quantifying the number of qubits and gates needed for implementation, both for our new SOS algorithm and for previously developed techniques for implementing MPS states \cite{schon2005sequential}. To assess and compare the quality of the several candidate states in hand, we develop a methodology that works based on their associated \textit{energy distributions}. These are projections of the candidate wavefunction on the eigenspectrum of the system Hamiltonian: while obtaining them exactly is more difficult than computing the overlap with the ground state, they can be approximated -- a task for which we propose new classical and quantum methods. Once the state quality is assessed and the chosen state is implemented, it can also be further refined with the use of a quantum filtering algorithm. On the basis of our analyses, we find that coarse QPE with post-selection -- that is, QPE performed with low precision -- generally outperforms other filtering methods. 
	
	The concept of the energy distribution has utility beyond state quality assessment. First, it suggests an alternative interpretation of quantum energy estimation algorithms -- not as a means of projecting onto the ground state, but as a way to improve classical estimates of the ground-state energy. Second, when QPE is performed, the energy distribution can help address the problem of the contribution of higher energy states towards low outcome values -- what we call the {\it leakage problem} in QPE. We show that this problem can be diagnosed when the energy distribution is at hand, and can furthermore be mitigated through quantum refining mentioned above. Finally, the energy distribution picture can be a guide towards potential quantum advantage: the amount of low-energy support of the initial state below a classical target energy estimate can be a proxy for the likelihood that quantum energy estimation algorithms can obtain lower energy estimates than the classical reference. With this in mind, we introduce the concept of \textit{Goldilocks problems}: energy estimation tasks where the initial state is neither too good (where classical methods are sufficient) nor too bad (where even quantum algorithms fail). Our energy distribution techniques can be used to search for such problems, which are candidates for quantum advantage, as we illustrate with numerical examples.
	
\begin{figure*}
		\centering
		\includegraphics[width=0.99\textwidth]{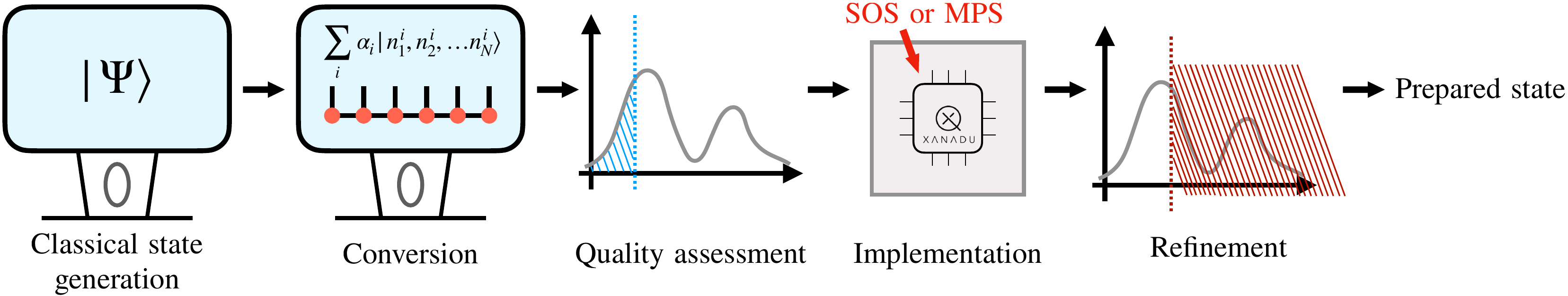}
		\caption{The initial state preparation algorithm consists of the following steps:
			classical search for a low energy state, conversion of the state into a standardized form, i.e.~SOS or MPS, quality assessment performed based on the energy distribution of the candidate state, implementation of the state on the quantum computer, and quantum refining. The end-to-end procedure results in a high quality state prepared on the quantum computer.}
		\label{fig:pipeline}
	\end{figure*}

	All of the subroutines discussed above combine to give an initial state preparation algorithm, which can be applied for quantum energy estimation in any quantum chemical system. The complete algorithm, illustrated in \cref{fig:pipeline} consists of the following steps:
	\begin{enumerate}
		\item Classical computation of a candidate initial state 
		\item Converting the candidate initial state to either SOS or MPS form
		\item Assessing the quality of different candidates through the energy distribution
		\item Implementing the resulting state into the quantum computer
		\item Quantum refining of the state with an energy filtering method
		\item With the implemented state, execute QPE or any other quantum energy estimation algorithm.
	\end{enumerate}
	Using our algorithm, we find that for nontrivial problems of interest in quantum chemistry such as estimating ground-state energies of iron-sulfur complexes, by improving state quality we can reduce the total algorithm cost by several orders of magnitude, compared to using a single product state. 
	
	The rest of the paper is structured as follows. In \cref{sec:classical-compute}, we briefly review the traditional quantum chemistry methods we use for calculating candidate initial states. In the following \cref{sec:quantum-implement}, we describe our state-of-the-art algorithm for implementing such initial states in their SOS form, as well as a separate way of implementing a state in the MPS form, providing resource estimations for both approaches. To assess and compare our candidate states, in \cref{sec:energy-distribution} we introduce the concept of the energy distribution, pioneering new methods to estimate the quality of candidate states. As a case study, we apply our new techniques to address the QPE leakage problem. Once a state is prepared, assessed and implemented, we can use a quantum algorithm to refine it by filtering out high-energy contributions, as we discuss in \cref{sec:quantum-refining}. Having described each step of our algorithm individually, in \cref{sec:numerics} we look at the entire pipeline, and showcase numerical experiments that demonstrate the state preparation algorithm for different molecules.
	
	\section{Obtaining classical descriptions of initial states}
	\label{sec:classical-compute}
	
	The initial state preparation algorithm starts from executing a traditional quantum chemistry method to generate a candidate wavefunction. It is likely that the choice of which method to run is highly situation-dependent, so we consider a wide variety of techniques, which we now briefly describe. We also review the concept of electronic correlation and its implications for choosing a quantum chemistry method for a given molecule. Experts are welcome to skip this section, but concepts from it will be employed throughout our paper.
	
	Throughout this paper, we work in the second quantized language and represent states with Fock occupation number vectors. By a Slater determinant we mean a many-particle product state built by distributing the available $n_e$ electrons over $2N$ single-particle spin-orbitals. We will use either spatial orbital occupation numbers $n_i = 0, \alpha, \beta, 2$ or spin-orbital occupation numbers $n_i = 0, 1$, typically ordered according to their Hartree-Fock energy (increasing from left to right): the choice of whether orbital or spin-orbital occupations are used will be clear from context. For example, a generic $S_z = 0$ Hartree-Fock state is written as $\ket{222...200...}$ in the first scheme and $\ket{1111...100...}$ in the second scheme, respectively. More broadly, a generic Slater determinant reads
	\begin{equation}
		\ket{\psi_{\text{Slater}}} = \ket{ n_1, n_2, \ldots, n_N} \text{ or } \ket{ n_1, n_2, \ldots, n_{2N}}.
	\end{equation}
	
	\subsection{Quantum chemistry methods for obtaining approximate ground states}
	
	The strategies we consider can be split into two groups. The first includes methods in the configuration interaction family, building the wavefunction as a superposition of Slater determinants (SOS). The second includes just one approach: representing the wavefunction using the matrix product state (MPS) ansatz, and variationally optimizing it over a series of sweeps using the density matrix renormalization group algorithm (DMRG)~\cite{white1992density, white1993density}. We focus first on methods based directly on the basis of Slater determinants. 
	
	The configuration interaction with single and double excitations (CISD) aims to prepare a wavefunction of the form
	\begin{equation}
		\ket{\psi_{\text{CISD}}} = \ket{\text{HF}} + \sum_i c_i \ket{S_i} + \sum_{i} d_i \ket{D_i},
	\end{equation}
	with singly and doubly excited determinants $S_i, D_i$ that parametrized by CI coefficients $c_i, d_i$. The coupled cluster with singles and doubles (CCSD) technique instead builds a wavefunction of the form
	\begin{equation}
		\ket{\psi_{\text{CCSD}}} = e^{\hat{T}_1 + \hat{T}_2} \ket{\text{HF}},
	\end{equation}
	where the excitation operators $\hat{T}_1, \hat{T}_2$ are single and double excitation operators parametrized by amplitudes $t^{(1,2)}$ that only connect occupied orbitals to virtual ones. Both methods are built on top of the Hartree-Fock reference state $\ket{\text{HF}}$ and are hence termed single-reference methods.  
	
	On the other hand, instead of restricting the many-body Hilbert space in terms of excitations, we can do this at the level of the single-particle basis: this is the nature of the complete active space configuration interaction (CASCI) method. With $N$ spatial orbitals and $1 \leq l < L \leq N$, we fix the first $l-1$ orbitals to be fully occupied and set the final $N-L-1$ orbitals to be unoccupied. Then the wavefunction takes the form
	\begin{multline}
		\label{eq:casci_wf}
		\ket{\psi_{\text{CASCI}}} = \\
		\sum_{ n_l, n_{l+1}, ... n_{L}} c_{n_l, n_{l+1}, ..., n_{L} } \ket{ 2, ... 2, n_l, n_{l+1}, ..., n_L, 0, ..., 0 },    
	\end{multline}
	parametrized by $c_{n_l, n_{l+1}, ..., n_{L}}$. CASCI executes exact diagonalization within the active space of orbitals $\{l, l+1, ..., L\}$. While this lets CASCI produce complex multireference wavefunctions that can have many determinants with similar weights, there are two limitations. First, the choice of active orbitals is widely acknowledged to be challenging~\cite{veryazov2011select} (although automated approaches \cite{sayfutyarova2017automated, stein2019autocas} are gaining popularity). Second, exact diagonalization scales too prohibitively to be useful in nontrivial molecules. The impact of frozen orbitals on the CASCI wavefunction could be partially taken into account with multireference perturbation theory (MRPT), which is applied on top of a CASCI calculation. The perturbative correction modifies the coefficients $c_{n_l, ... n_L}$ of the CASCI wavefunction, typically using standard second-order (Moller-Plesset) perturbation theory. In practice, MRPT is usually much better at improving the energy estimate than at improving the wavefunction: largely used for recovering dynamic correlation energy, it is not capable of adding multireference character from the non-active space states. 
	
	Selective configuration interaction (SCI) methods are inspired by the idea that for many wavefunctions of interest, written in the full basis as
	\begin{equation}
		\ket{\psi_{\text{SCI}}} = \sum_{n_1, n_{2}, ... n_{N}} c_{n_1, n_{2}, ..., n_{N} } \ket{ n_1, n_{2}, ..., n_N },    
	\end{equation}
	most coefficients $c_{n_1, n_{2}, ..., n_{N} }$ vanish. The goal is to identify an efficient way of searching for these non-zero coefficients \cite{tubman2016deterministic, schriber2016communication, holmes2016heat, harrison1991approximating, tubman2020modern, bender1969studies, huron1973iterative}. We focus on the recently developed semistochastic heat-bath configuration interaction (SHCI) \cite{sharma2017semistochastic, holmes2016heat}, which employs the relatively simple criterion $\max_{i}(H_{ki} c_i) > \epsilon_1$: here $H_{ki}$ is the Hamiltonian matrix element between determinants in the variational basis $c_i$ and a candidate external determinant $c_k$, and $\epsilon_1$ a user-chosen cutoff.
	
	Finally, we turn to the second group and summarize the DMRG approach for obtaining wavefuctions in MPS form. DMRG has proven to be a reliable, robust, and efficient method for constructing approximate ground states for a wide variety of molecules \cite{white1999ab, marti2008density, kurashige2013entangled, sharma2014low, olivares2015ab, von2021quantum, larsson2022chromium, lee2023evaluating}. An MPS can be seen as an efficient way of factorizing the general $N$-tensor coefficient $c_{n_1, ..., n_N}$ of a Slater determinant series into a product of matrices, whose internal dimension is limited by the bond dimension $\chi$ \cite{chan2008introduction}. The MPS wavefunction can be written as
	\begin{align}\label{eq:mps_form}
		\ket{\psi_{\text{MPS}}} = \sum_{ \mathclap{ \substack{\alpha_1,\ldots,\alpha_{N-1} \\ n_1, \ldots, n_N } } } & A^{n_1}_{1; \, \alpha_1} \, A^{ n_2}_{2; \, \alpha_1 \alpha_2} \ldots A^{ n_N}_{N; \, \alpha_{N-1}} \ket{n_1 , n_2 , \ldots , n_N}, \nonumber \\
		& n_i \in \{0, \alpha, \beta, 2 \}, \quad \alpha_i \leq \chi.
	\end{align}
	This factorization scales polynomially with system size for a fixed bond dimension~\cite{chan2011density,schollwock2011density}. 
	Combined with the DMRG algorithm, it gives a wavefunction-based variational approach that provably converges to the exact solution in the limit $\chi \rightarrow \infty$. 
	To apply the DMRG method to molecules, which exhibit inherently nonlocal interactions between molecular orbitals and do not resemble the spin chains that DMRG was originally developed for, orbitals must be arranged along a one-dimensional chain. Ideally, the arrangement is such that it minimizes the amount of long-range nonzero molecular integrals. The standard choice is to arrange the molecular orbitals according to their Hartree-Fock energy; more sophisticated reordering schemes are also considered~\cite{olivares2015ab}. 
	
	The diversity of wavefunction forms resulting from different methods means that in practice it is difficult to compare and evaluate them. To do the comparison, we will use the publicly available software package Overlapper \cite{overlapper}.
	On the other hand, this variety suggests that there are trade-offs that could be exploited depending on the molecule being considered. The strengths and weaknesses of different approaches are largely determined by the amount and type of correlation present in the system, as discuss next.
	
	\subsection{Electronic correlations}
	\label{subsec:correlations}
	
	Correlation energy is the portion of total system energy that is not accounted for by the Hartree-Fock ansatz. It is due to Coulomb interactions between electrons. 
	
	Strong correlation necessitates the use of additional determinants in the many-body wavefunction for an accurate description. Thus, depending on the amount of correlation and its type, different methods might be preferable.
	
	Correlation energy is usually partitioned into static and dynamic types. 
	Dynamic correlation is a consequence of electrons avoiding each other due to Coulomb repulsion. When electrons approach each other in real space -- as when sharing a spatial orbital -- their wavefunctions acquire non-analytic cusps due to the Coulomb potential divergence. Resolving these cusps requires a large basis set and many determinants. Dynamic correlation is thus associated with the many-body wavefunction being described by one dominant determinant with a large weight, together with many small-weight contributions. While the weights are small, the determinants are numerous, resulting in large energy errors. Fast single-reference methods such as CISD and CCSD are usually capable of recovering dynamic correlation energy, as they are able to work with a larger basis set and their single-reference nature is conducive to the task. 
	On the other hand, scale limitations of CASCI, DMRG and SHCI make them worse at recovering this type of correlation. 
	
	By contrast, static correlation arises in the presence of nearly-degenerate eigenstates: multiple determinants with roughly similar weights are needed for an accurate description of the ground state wavefunction. 
	Typical situations where static correlation arises are nonequilibrium geometries, low-spin states of open-shell molecules (spin state degeneracy), excited states and molecules containing transition metal atoms (due to high degeneracy of $d$-type orbitals). 
	In such situations a single reference, such as the Hartree-Fock state or CISD and CCSD, will not be a good starting point -- a multireference method such as CASCI, DMRG or SHCI is needed. 
	In practice, CASCI is strongly limited by the active space size; and while MRPT improves the CASCI energy appreciably through adding dynamic correlation, it provides only minor improvements to the wavefunction itself, which is the object of interest in quantum algorithms. This leaves DMRG and SHCI as the leading contenders. DMRG can boast polynomially efficient representation of even strongly multireference states and relatively straightforward convergence with bond dimension; by contrast, SHCI's lower computational demands make it easier to run calculations for larger spaces, without sacrificing accuracy even for correlated systems such as the chromium dimer~\cite{larsson2022chromium}.
	
	\section{Efficient ansatz implementation on quantum computers}
	\label{sec:quantum-implement}
	In \cref{sec:classical-compute}, we detailed how approximate ground states can be found using classical computational methods and expressed in two standardized forms: SOS and MPS. In this section, we detail how SOS and MPS states can be implemented on a quantum computer and estimate the number of qubits and gates required for these tasks. Typically the cost of encoding classical states is lower than the cost of the energy estimation algorithm, even for sophisticated states with many Slater determinants or large bond dimension. This results in considerable runtime reductions of the full algorithm by lowering the number of repetitions needed to achieve a target accuracy, while incurring only small increases in the cost of each independent run.
	
	\subsection{Sum of Slater determinants (SOS)}
	We employ the formalism of second quantization, but the implementation method is general and can also be used for algorithms employing a first quantization representation \cite{zini2023quantum}.
	
	The goal is to prepare the normalized state 
	\begin{equation}\label{eq:SOS_state}
		\ket{\psi} =  \sum_{i=1}^{D} \alpha_i\ket{\nu_i},
	\end{equation}
	where $\alpha_i$ are the given amplitudes and $\ket{\nu_i} = \ket{n_{i,1}n_{i,2}\cdots n_{i,2N}}$ are states of $2N$ qubits. The bits $n_{i,j}$ denote the occupation number of spin-orbital $j$ for the $i$-th Slater determinant $\ket{\nu_i}$. We use $N$ to denote the total number of spatial orbitals, each supporting one spin-up and one spin-down particle. 
	
	We are interested in cases where the number of Slater determinants $D$ is much smaller than  $2^{2N}$, which is the case in practice. Therefore, the problem is to prepare a summation of relatively few basis states picked from a very large Hilbert space. The gate cost of the algorithm will be measured in terms of the number of non-Clifford Toffoli gates, which is a standard complexity measure used in the literature. On a fault-tolerant architecture, Toffolis require orders of magnitude more qubit-seconds and physical qubits due to the need for magic state distillation to implement them. Since our algorithm is explicitly meant for state preparation on fault-tolerant quantum computers, we use Toffoli gate count (and not CNOT gate count) to measure the cost.
	
	There has been previous work in this direction. Ref.~\cite{tubman2018postponing} proposes an iterative generation of the state using $(2N-1)(D-1)$ Toffoli gates and $2N-1$ ancilla qubits (when $D>1$). Other algorithms include \cite{ortiz2001quantum,babbush2015chemical}, but have Toffoli complexity even higher than $O(ND)$, or potentially exponential in $N$ \cite{wang2009efficient}. Instead, we present an algorithm with asymptotic runtime of $O(D \log D)$, where $\log$ is in base two throughout this manuscript. 
	This is a considerable improvement since the number of Slaters $D$ is at most the full space $2^N$, and so, it is often the case that $\log D \ll N$: \cref{fig:modji-vs-tubman} makes this comparison explicit for a few different model systems and molecules studied in this paper and elsewhere. Notice that the advantage is even more explicit for larger systems; for example, with $N\sim 400$ spin-orbitals, our algorithm is an order of magnitude more efficient in Toffoli cost as long as $D < 2^{40} \sim 10^{12}$.
	
	We aim to prepare the state in \cref{eq:SOS_state},
	where $\ket{\nu_i}$ represents the occupation bitstrings of length $2N$ mentioned above. Our main technical result is a mapping from the bitstrings $\nu_i$ of $2N$ bits, that identify each Slater determinant, to more compact and unique bitstrings $b_i$ of only $O(\log D)$ bits. The following lemma, formalizes the compression scheme. We will assume that $D$ is a power of two for convenience; otherwise, $\log D$ should be replaced with its ceiling.
	
	\begin{lem}\label{lem:unique_signature}
		Given as input a set $\{\nu_i\}$ of bitstrings representing unique Slater determinants, there is a classical algorithm with complexity $O(tD^2)$, where $t \le \min(2N,D) - 2\log D +1$, that outputs substrings $\tilde{\nu}_i$ of $\nu_i$ and $2 \log D-1$ bitstrings $u_k$ of length $O(D)$, such that the bitstrings $b_i := ( u_1 \cdot \tilde{\nu}_i, \ldots , u_{2\log D-1} \cdot \tilde{\nu}_i)$ presented as column vectors in
		\begin{align}\label{eq:U_V_B}
			\begin{pmatrix}
				u_1^T \\
				\vdots \\
				u_{2\log D-1}^T
			\end{pmatrix}
			&\begin{pmatrix}
				\tilde{\nu}_1 & \cdots & \tilde{\nu}_D
			\end{pmatrix} =\begin{pmatrix}
				b_1 & \cdots & b_D
			\end{pmatrix},
		\end{align}
		are mutually distinct, i.e., $b_i \neq b_j$ for all $i$ and $j$.
	\end{lem}
	The proof can be found in \cref{appsec:proof_of_compression_lemma}. The algorithm to implement the SOS state in \cref{eq:SOS_state} is described below. We employ three registers: the system register, where we wish to prepare the desired state, and two ancilla registers: an enumeration register with $\log D$ qubits and an identification register with $2\log D-1$ qubits. \\
	
	\textbf{Quantum algorithm for encoding SOS states}
	
\begin{figure}[!t]
		\centering
		\includegraphics[width=0.9\columnwidth]{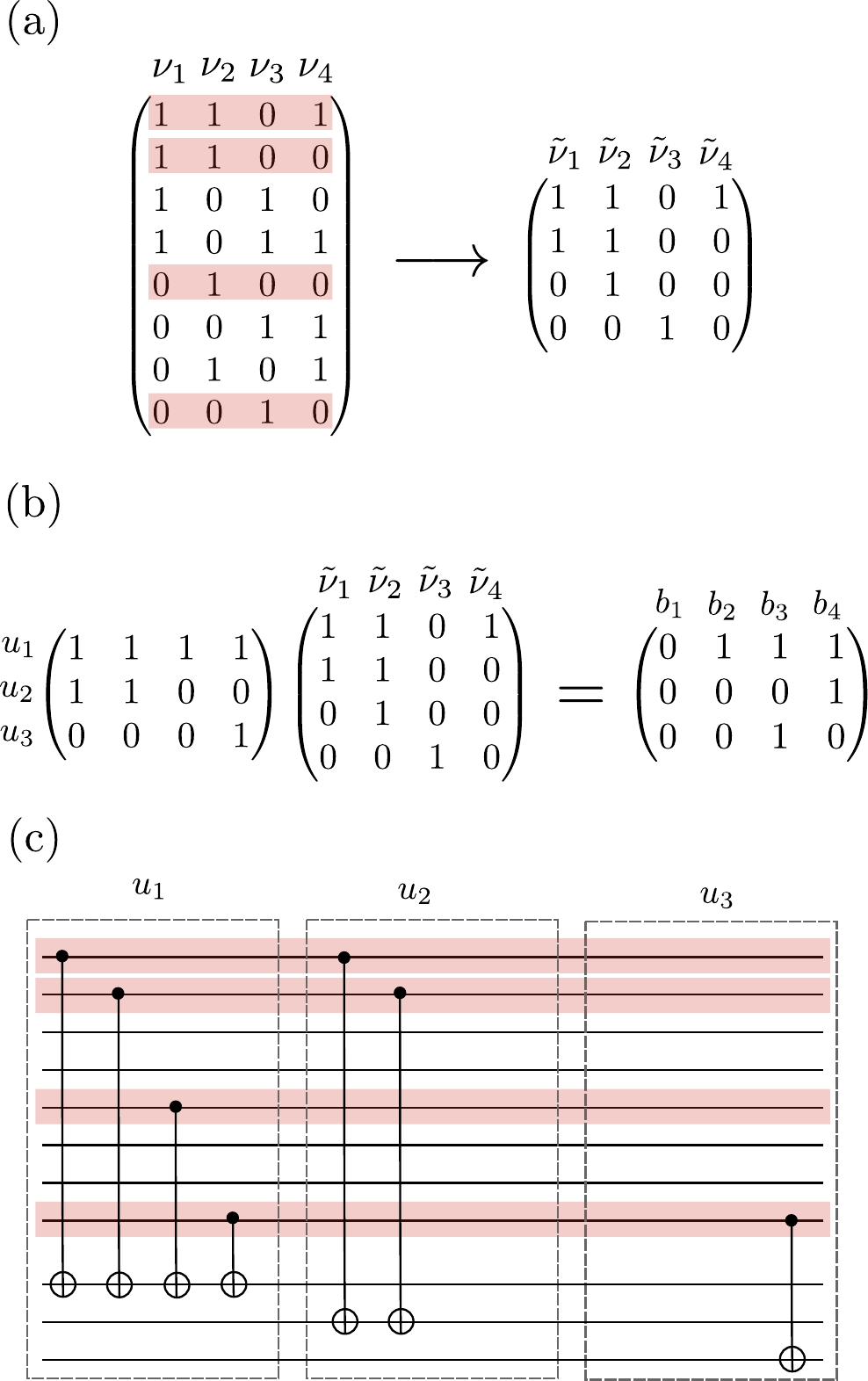}
		\caption{Schematic representation of important steps in the SOS encoding algorithm. We consider an example with $2N=8$ orbitals and $D=4$ Slater determinants, for which the identification strings require $2 \log D -1=3$ bits. (a) Matrix of Slater determinant strings $\nu_i$. By selecting only rows 1,2, 5, and 8, we can construct substrings $\tilde{\nu}_i$ that form a matrix of full rank. (b) Using the result from \cref{lem:unique_signature}, we construct bitstrings $u_i$ that form a linear map transforming the substrings $\tilde{\nu}_i$ to the identification bitstrings $b_i$. (c) The encoding quantum algorithm applies a series of CNOT operations for each $u_i$, acting only on qubits 1, 2, 5 and 8 in the system register, in accordance to the choice of bitstrings. These are responsible for setting every individual qubit in the identification register to the desired value. }
		\label{fig:sos-algorithm}
	\end{figure}

\begin{figure}[t]
		\centering
		\includegraphics[width=0.5\textwidth]{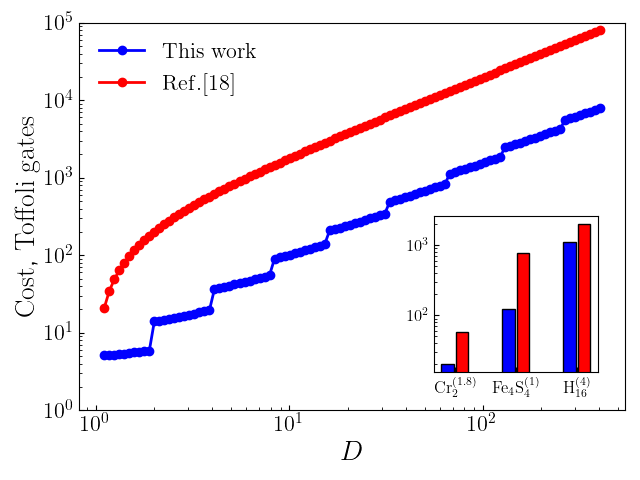}
		\caption{Comparing the cost of implementing an SOS wavefunction with $D$ determinants between the algorithm proposed here and that of Ref. \cite{tubman2018postponing} for a hypothetical system with $N = 100$ spatial orbitals. Inset: comparing the cost of preparing an SOS wavefunction of quality 0.2, as measured through overlap with the highest-fidelity reference, for specific molecules. In the inset, the superscript on the molecule formula indicates how much the bond is stretched relative to equilibrium. Note that system size $N$ varies between the molecules: $10$ for Cr$_2$, $32$ for Fe$_4$S$_4$, and $16$ for H$_{16}$. The number of determinants required to achieve quality of 0.2 also varies.}
		\label{fig:modji-vs-tubman}
	\end{figure}

	\begin{enumerate}
		\item Prepare the state
		\begin{align}\label{eq:superposition_enumerator}
			\sum_{i=1}^{D} \alpha_i \ket{0}\ket{i}\ket{0},
		\end{align}
		in the enumeration register using the Quantum Read-Only Memory (QROM) state preparation method in~\cite{low2018trading}. 
		\item  Use a QROM oracle $O$ of Toffoli cost $D$ as in Ref.~\cite{low2018trading} that implements the transformation 
		\begin{align}\label{eq:first_qrom_sys_register}
			O \ket{0}\ket{i}\ket{0} = \ket{\nu_i}\ket{i}\ket{0}.
		\end{align}
		This results in the state
		\begin{align}
			\sum_{i=1}^{D} \alpha_i\ket{\nu_i}\ket{i}\ket{0}.
		\end{align}
		\item Using the output bitstrings $u_k$ from \cref{lem:unique_signature}, do the following. If the $j$-th bit of $u_1$ is equal to 1, apply a CNOT gate between the system register and the identification register. The CNOT is controlled on the $j$-th qubit of $\ket{\nu_i}$. This results in the state
		\begin{align}
			&\sum_{i=1}^{D} \alpha_i\ket{\nu_i}\ket{i}(\ket{u_1\cdot \tilde{\nu}_i}\ket{0})\\
			=&\sum_{i=1}^{D} \alpha_i\ket{\nu_i}\ket{i}(\ket{b_{i,1}}\ket{0}).
		\end{align}
		Notice the bits in $u_1$ are matched with the bits of the substring $\tilde{\nu}_i$ obtained from \cref{lem:unique_signature}.
		\item Repeat the above step for all $u_k$. This results in the state
		\begin{align}
			\sum_{i=1}^{D} \alpha_i\ket{\nu_i}\ket{i}\ket{b_i},
		\end{align}
		which now contains the unique compact identifier $b_i$ for the Slater determinant $\ket{\nu_i}$.
		\item Using multi-controlled operations, apply the transformation $\ket{i}\rightarrow \ket{0}$ to the first ancilla register conditioned on each unique $b_i$. This results in the state
		\begin{equation}
			\sum_{i=1}^{D} \alpha_i\ket{\nu_i}\ket{0}\ket{b_i}.
		\end{equation}
		\item The final step is to uncompute the sequence of CNOTs used to prepare the state $\ket{b_i}$ in steps 3 and 4. This leads to the final output
		\begin{equation}
			\sum_{i=1}^{D} \alpha_i\ket{\nu_i}\ket{0}\ket{0},
		\end{equation}
		which contains the target state in the system register disentangled form all other registers, as desired. 
	\end{enumerate}  
	Important steps of the algorithm are illustrated in \cref{fig:sos-algorithm}.
	
	Taking into account the first two usages of QROM for preparing $\sum_{i=1}^{D} \alpha_i \ket{i}$  and in \cref{eq:first_qrom_sys_register}, the overall Toffoli cost is dominated by 
	\begin{align}
		&(2\log D-2)D + 2^{\log D+1} + D < \nonumber \\
		&(2\log D+3)D = O(D \log D).
	\end{align}
	
	The overall additional qubit cost due to use of ancillas is  
	\begin{align}
		4\log D-3+\log D = 5\log D-3 = O(\log D).
	\end{align}
	
	One could trade off Toffolis with qubits, within the same volume cost of $O((\log D)^2D)$. In most cases, this means using the \ssdirty variant of QROM \cite{low2018trading}, also called QROAM in~\cite{berry2019qubitization}. Importantly, this variant allows using uninitialized qubits for this trade-off. The trade-off leads to a Toffoli cost of $\min(2\sqrt{32ND},D) + (7\log D+2\sqrt{32\log D})\sqrt{D}$, a clean qubit cost of $(2\log D-1)\sqrt{D}$, and an additional uninitialized qubit cost of $\sqrt{32ND}$. We explain this in details in \cref{appsec:trading_off_tof_w_qubits}, along with comments on how one could lower the expected Toffoli cost by combining our strategy with \cite{tubman2018postponing}.
	
	\subsection{Matrix-product state (MPS)}\label{sec:mps_implementation_main}
	Here we discuss how to implement initial states expressed in the MPS form. While it is always possible to transform an MPS into an SOS formulation, direct implementation of an MPS can be beneficial in certain cases. We consider mainly the method introduced in Ref.~\cite{schon2005sequential} and perform an estimation of the total Toffoli cost. We note that there are many variations of this technique considered in the literature \cite{ran2020encoding,rudolph2022decomposition,dov2022approximate} and also newer versions requiring lower-depth circuits~\cite{malz2023preparation} for short-range correlated MPSs. 
	
	First, we give a quick review of the method in Ref.~\cite{schon2005sequential}. We start with the MPS form shown in~\cref{eq:mps_form} and use standard graphical notation for representing the MPS. We denote its tensors as 
	\begin{equation}
		A^{n_j}_{j; \ \alpha_{j-1} \alpha_j} = 
		\begin{tikzpicture}[baseline=1ex]
			\draw[black, fill=red!10,  thick] (0,0) rectangle (0.5,0.5) node[pos=.5] {$A_j$};
			\draw[thick] (0.,0.25) -- (-0.25,0.25) ;
			\path (-0.65,0.25) node {$\alpha_{j-1}$};
			\draw[thick] (0.25,0.5) -- (0.25,0.75) ;
			\path (0.5,0.8) node {$n_j$};
			\draw[thick] (0.5,0.25) -- (0.75,0.25) ;
			\path (1.,0.25) node {$\alpha_j$};
		\end{tikzpicture}    
		.
	\end{equation}
	The physical index $n_j$ runs over $d$ values, where $d$ is the local Hilbert space dimension. The auxiliary indices $\alpha_{j}$ run over $\chi_j$ values, where $\chi_j$ is called the bond dimension, which generally may be different for each index $j$.
	For the implementation, the MPS is turned into its left-canonical form:
	\begin{equation}
		\begin{tikzpicture}
			\draw[black, fill=red!10,  thick] (-.75,0) rectangle (-.25,0.5) node[pos=.5] {$A_1$};
			\draw[thick] (-0.5,0.5) -- (-0.5,0.75) ;
			\path (-0.25,0.8) node {$n_1$};
			\draw[thick] (0.,0.25) -- (-0.25,0.25) ;
			\draw[black, fill=red!10,  thick] (0,0) rectangle (0.5,0.5) node[pos=.5] {$A_2$};
			\draw[thick] (0.25,0.5) -- (0.25,0.75) ;
			\path (0.5,0.8) node {$n_2$};
			\draw[thick] (0.5,0.25) -- (0.75,0.25) ;
			\draw[black, fill=red!10,  thick] (0.75,0) rectangle (1.25,0.5) node[pos=.5] {$A_3$};
			\draw[thick] (1.,0.5) -- (1.,0.75) ;
			\path (1.25,0.8) node {$n_3$};
			\draw[thick] (1.25,0.25) -- (1.5,0.25) ;
			\path (2.15,.25) node {$\ldots$};
			\draw[thick] (2.75,0.25) -- (3.,0.25) ;
			\draw[black, fill=red!10,  thick] (3.,0) rectangle (3.5,0.5) node[pos=.5] {$A_{N}$};
			\draw[thick] (3.25,0.5) -- (3.25,0.75) ;
			\path (3.5,0.8) node {$n_{N}$};
		\end{tikzpicture}  .
	\end{equation}
	It means that for all $j>1$, we have:
	\begin{equation}\label{eq:left_ortho}
		\sum_{\alpha_j,n_j} A_{j; \ \alpha_{j-1}\alpha_j}^{n_j} \left(A_{j; \ \alpha_{j-1}'\alpha_j}^{n_j}\right)^* = \delta_{\alpha_{j-1} \alpha_{j-1}'},
	\end{equation}
	or diagrammatically:
	\begin{equation}
		\begin{tikzpicture}[baseline=4ex]
			\draw[thick] (0.5,0.25) -- (0.75,0.25) ;
			\draw[black, fill=red!10,  thick] (0.75,0) rectangle (1.25,0.5) node[pos=.5] {$A_j$};
			\draw[thick] (1.,0.5) -- (1.,1) ;
			\draw[thick] (1.25,0.25) -- (1.5,0.25) ;
			\draw[thick] (0.5,1.25) -- (0.75,1.25) ;
			\draw[black, fill=red!10,  thick] (0.75,1.) rectangle (1.25,1.5) node[pos=.5] {$A_j$};
			\draw[thick] (1.25,1.25) -- (1.5,1.25) ;
			\draw[thick] (1.5,0.25) -- (1.5,1.25) ;
		\end{tikzpicture}
		\ \ =  \ \
		\begin{tikzpicture}[baseline=4ex]
			\draw[thick] (1.25,0.25) -- (1.5,0.25) ;
			\draw[thick] (1.25,1.25) -- (1.5,1.25) ;
			\draw[thick] (1.5,0.25) -- (1.5,1.25) ;
		\end{tikzpicture}\ .
	\end{equation}
	This is in general possible using singular value decomposition of tensors \cite{schollwock2011density}.
	A note regarding notation: in the above equations and everywhere else, the summation over left and right auxiliary indices of the leftmost and the rightmost tensors, respectively, can simply be dropped. This is equivalent to setting $\chi_0,\chi_{N+1}=0$.
	
	The implementation works by first observing that the above tensors of the MPS, owing to the left-canonical form, can be directly used to define unitaries, which we denote as $G$, that are used in a quantum circuit for preparing the MPS:
	\begin{equation}\label{eq:G_def_tensors_grpah}
		\begin{aligned}
			G[j]_{\alpha_{j}n_j , \alpha_{j-1} 0} &= A_{j; \ \alpha_{j-1}\alpha_j}^{n_j}, \\
			\begin{tikzpicture}[baseline=1ex]
				\draw[black, fill=red!10,  thick] (0,0) rectangle (0.5,0.5) node[pos=.5] {$A_j$};
				\draw[thick] (0.,0.25) -- (-0.25,0.25) ;
				\path (-0.65,0.25) node {$\alpha_{j-1}$};
				\draw[thick] (0.25,0.5) -- (0.25,0.75) ;
				\path (0.5,0.8) node {$n_j$};
				\draw[thick] (0.5,0.25) -- (0.75,0.25) ;
				\path (1.,0.25) node {$\alpha_j$};
			\end{tikzpicture}   & \ \to \
			\begin{tikzpicture}[baseline=1ex]
				\draw[black, fill=red!10,  thick] (0,0) rectangle (0.8,0.8) node[pos=.5] {$G[j]$};
				\draw[thick] (-0.25,0.2) -- (0.,0.2) ;
				\path (-0.7,0.1) node {$\alpha_{j-1}$};
				\draw[thick] (-0.25,0.6) -- (0.,0.6) ;
				\path (-0.55,.6) node {$0$};
				\draw[thick] (0.8,0.2) -- (1.05,0.2) ;
				\path (1.3,0.1) node {$\alpha_j$};
				\draw[thick] (0.8,0.6) -- (1.05,0.6) ;
				\path (1.3,0.6) node {$n_j$};
			\end{tikzpicture}
			.
		\end{aligned}
	\end{equation}
	Each unitary acts on a $d$-level system composed of $\lceil \log(d)\rceil $ qubits as well as $\lceil\log \chi\rceil$ ancillae, where we remind the reader that $\log$ is in base two throughout this manuscript. For example, for fermionic systems, $d=4$ and two qubits are required for each spatial orbital of the system.
	Furthermore, the incoming physical index for the unitary is set to be equal to 0 and thus the above relation does not specify all the elements of $G[j]$. This is fine as long as the rest of the elements are chosen so that $G[j]$ remains unitary. 
	
\begin{figure}[!t]
		\centering
		\includegraphics[width=0.48\textwidth]{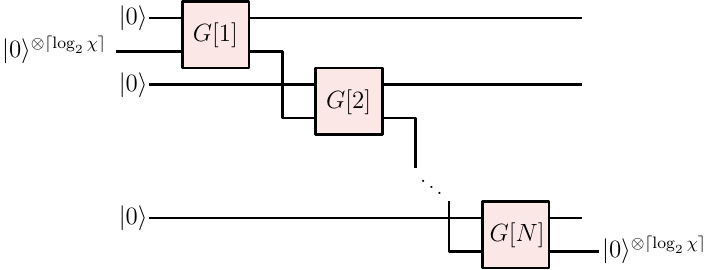}
		\caption{Circuit for MPS implementation. Since the bond dimension can change as the circuit traverses the system, one can start with a number of ancillae equal to $\lceil \log \chi_{\text{max}} \rceil$ and use more or less of the available anciallae as the bond dimension dictates.}
		\label{fig:mps_implementation_circuit}
	\end{figure}

	A quantum circuit that implements the desired MPS and works with these unitaries is shown in \cref{fig:mps_implementation_circuit}. Note that an auxiliary register of a size, which we have denoted collectively as $\lceil \log \chi \rceil$, is required to reconstruct the MPS and that we are schematically moving it around to act with unitaries on this register and different qudits in the system. 
	Also, note that for the first and the last unitaries the input and output auxiliary registers also have value $0$.

	For the above circuit, we need to synthesize the unitaries $G[j]$, for which we use the method in Ref.~\cite{low2018trading}. 
	Details of how the synthesis can be performed is discussed in \cref{app:mps_implementation_details}. There, it is shown that each $G[j]$ imposes a Toffoli cost of:
	\begin{equation}
		\chi_{j-1} \left[ 8\chi_j d + (b+1) \log(\chi_j d) \right],
	\end{equation}
	where $b$ is the number of digits for storing the angle for implementing single qubit rotations required in the synthesis process. 
	
	Assuming a number $N$ of qudits in the physical system, the total cost will be the sum of the above over all $j$; asymptotically and with using $\chi$ selectively for all bond dimensions, the dominant Toffoli cost can be written as $O(N \chi^2)$.
	Using trade-off schemes of \cite{low2018trading}, we can show that it is also possible to use $O(N \chi^{3/2})$ Toffolis for implementation of the MPS.

	\subsection{Discussion of the implementation methods}
	
	As with many questions concerning initial state preparation, the choice of SOS vs MPS is highly context-dependent. The polynomial scaling (with system size) of the number of parameters in an MPS may provide it with a decisive advantage for strongly multireference systems; while in single-reference molecules (or ones not being highly multireference), the reduced cost of implementing the SOS form might be preferable. A concrete comparison between these two methods is done in \cref{sec:H_chains_numerics}. Either way, these two forms are inter-convertible, and we have outlined advanced quantum algorithms for implementing states of either form. This includes a novel method for SOS states with lower asymptotic cost than the previous state of the art. Crucially, the cost of implementing a classical state on a quantum computer is typically a lot lower than that of the full energy estimation algorithm. This means there is a large budget available for implementing sophisticated states with many Slater determinants or large enough bond dimension that better approximate the true ground state compared to simpler approaches like the Hartree-Fock state. This can lead to considerable runtime reductions for the entire quantum algorithm.

	\section{Energy distribution of the initial state}
	\label{sec:energy-distribution}
	Using the overlap with the ground state as a way to assess the quality of an initial state is a challenging task, especially in strongly correlated systems. This is largely because the true ground state is generally not known: after all, this is the problem we are attempting to solve.  Instead, we propose a new way to assess state quality through the use of the state's associated \textit{energy distribution}. In this section, we first define energy distributions in precise terms. We then discuss how the energy distribution picture can change our view of performing quantum phase estimation (QPE). Our main technical contribution is a description of methods for approximating energy distributions, as well as formalizing how the energy distribution picture can be used to predict statistics of QPE outcomes. Finally, we detail how the QPE leakage problem can be seen through the energy distribution of the initial state.
	
	We define the energy distribution of the state $\ket{\psi}$ with respect to the Hamiltonian $H$ as:
	\begin{equation}\label{eq:def_energy_dist}
		P(E) = \sum_n |\langle E_n| \psi\rangle|^2 \, f_\eta(E_n-E),
	\end{equation}
	where $E_n,\ket{E_n}$ are respectively eigenvalues and eigenstates of $H$, and $f_{\eta}$ is a kernel, for example Gaussian or Lorentzian, with width $\eta$, a copy of which is centered at each of the eigenvalues. The limit of $\eta \to 0$ corresponds to a discrete distribution --- the actual distribution of the state's overlaps with the Hamiltonian spectrum. With $\eta \neq 0$, each energy level is broadened and the result is a continuous distribution. In practice, it is hard to access the true $\eta=0$ distribution, but much can be inferred about state quality from the broadened distribution. 
	
	We use the energy distribution in a number of applications. Most importantly, we formulate a simple criterion based on the energy distribution for assessing the quality of initial states. Suppose we are given a number of candidate states with similar energies, i.e., with similar expectation values of $H$. We can compare the quality of the states by focusing on the left-side tails of the states' energy distributions. Intuitively, whichever state has more weight extended to lower energies is a better candidate, as it provides higher probability for obtaining a low-energy estimate. We will make this statement more precise in \cref{sec:lowest_outcome}.

	\subsection{Quantum phase estimation through the lens of energy distributions}
	In many of the quantum routines for quantum energy estimation, the energy distribution has a very close relation with the distribution of outcomes: for QPE, for example, the outcomes are roughly sampled from the energy distribution of the initial state. More precisely, in a QPE measurement with $k$ phase digits, the probability of an integer outcome $x_m$ (that can be interpreted as an estimated energy of $2^{-k} x_m$) in the phase register reads \cite{nielsen2010quantum}:
	\begin{equation}\label{eq:QPE_outcome_state_overlaps}
		\sum_{n} |\langle E_n|\psi\rangle|^2  \ \ \frac{1}{2^{2k}} \left(\frac{\sin^2(\pi 2^k E_n)}{ \sin^2(\pi [E_n - x_m/2^k ])} \right).
	\end{equation}
	Notice that in the above, we have assumed a normalization of the Hamiltonian such that $0 \leq E_n \leq 1$ for all $n$. We similarly assume that the integer outcome $x_m$ satisfies $0 \leq x_m < 2^k$. The discrete QPE kernel $\frac{1}{2^{2k}} \left(\frac{\sin^2(\pi 2^k E_n)}{ \sin^2(\pi [E_n - x_m/2^k ])} \right)$ which appears in the above relation, broadens each energy level and is maximized when the integer $x_m$ is closest to $2^k E_n$ for each level; thus, performing QPE can be thought of as sampling from a discrete distribution, which is obtained by spreading the weight of each energy level by the discrete QPE kernel. As the number of digits $k$ increases, the sampling gets closer to sampling from the actual underlying distribution.

	The traditional viewpoint for QPE is that the algorithm must be repeated enough times so that there is a high probability of sampling the ground-state energy (up to the allowed precision). This perspective tacitly implies that all samples other than the ground-state energy should be discarded as useless. 
	
\begin{figure}
		\centering
		\includegraphics[width=0.4\textwidth]{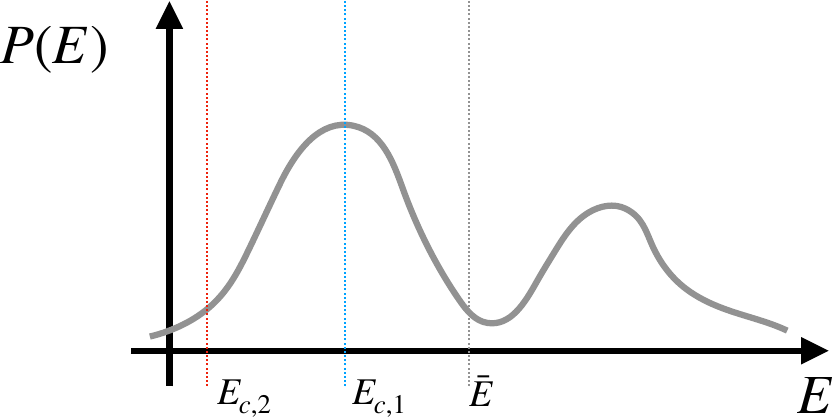}
		\caption{Schematic depiction of an energy distribution for a particular initial state, illustrating the possibility of improvement over classically found energy estimates. If the best classical energy estimate is $E_{c,1}$, there is a high chance of obtaining better quantum estimates using quantum routines e.g.~QPE. However, if the best classical estimate is $E_{c,2}$, the chance is quite slim. $\bar{E}$, the energy of the implemented state is also shown, but its value is irrelevant for predicting the likelihood of obtaining better energy estimates.
			Note that in practice, one must ensure that the weight to the left of each of these energy estimates is not due to the broadening of higher energy weight; such broadening ($\eta$ in \cref{eq:def_energy_dist}) is inevitable in any actual calculation, but one can examine the behavior of the tails as $\eta$ is varied to determine whether the weight in the tail is due to such broadening or not.}
		\label{fig:improvement_over_classical}
	\end{figure}

	Instead, we recognize that the goal of any algorithm, whether classical or quantum, is to provide the best possible energy estimates. Ideally this is precisely the ground-state energy, but that may be too ambitious in practice, especially for large systems with strong electronic correlations. The energy distribution picture presents an alternative where QPE is viewed as a method to \textit{improve} the energy estimate associated with the initial state. Intuitively, since the average of the energy distribution is precisely the classical estimate, QPE can equally sample energies that are higher or lower; even few repetitions can thus lead to better estimates. Generating a larger number of samples increases the probability of observing more dramatic improvements, with the ultimate goal of obtaining precisely the ground-state energy. Employing quantum algorithms is advantageous whenever there is a sizeable probability of obtaining an energy estimate that is lower than any classical method, including more powerful ones than those used for the initial state. This concept is illustrated in \cref{fig:improvement_over_classical}. 

	The energy distribution picture is useful beyond helping re-interpret QPE. The QPE kernel discussed above has long algebraic tails on the two sides of each energy level: as a result, it is in general possible that an outcome indicating a false low energy is obtained, that was actually in the long tail of much higher energy levels. Note that this is contrary to the occurrence of an outcome due to the actual weight in its vicinity; we call this phenomenon the QPE leakage problem. In fact, in general it is possible for the observed outcome to lie below the ground-state energy of the Hamiltonian, rendering it unphysical.  We will discuss the problem in detail along with ways it can be diagnosed and avoided -- by employing the energy distribution -- in \cref{sec:leakage_intro}.

	\subsection{Approximating energy distributions}
	
	We provide three different methods for approximating the energy distribution with respect to a Hamiltonian $H$ for an initial state $\ket{\psi}$. Two of them are classical methods and one is quantum. 
	
	\subsubsection{Series expansion}
	
	This method employs moments of energy (expectation values of powers of $H$) to obtain a series expansion for the energy distribution. 
	We consider the Edgeworth series and the Gram-Charlier series, which both approximate a distribution as a Gaussian multiplied by different orders of the Hermite polynomials. 
	The coefficients in the series can be written in terms of the moments of the distribution, i.e., the expectation values of powers of the Hamiltonian $\bra{\psi} H^n \ket{\psi}:=\langle E^n \rangle$. 
	The lowest order approximation is a Gaussian distribution with a variance proportional to that of the initial state, namely $\langle E^2 \rangle - \langle E \rangle^2$.
	This method works best for distributions that are nearly Gaussian.
	
	The Edgeworth and Gram-Charlier series have identical terms: the only difference is that the terms in an Edgeworth series are arranged in a way that the series constitutes a true asymptotic series \cite{blinnikov1998expansions}. 
	The Gram-Charlier series expansion for the energy distribution $P(E)$ can be written as
	\begin{equation}\label{eq:def_gram_charlier}
		\tilde{P}(E) = \frac{\exp(-E^2/2)}{\sqrt{2\pi}} \left[  1+\sum_{n=3}^{\infty} (-1)^n c_n \, \He_n(E)  \right],
	\end{equation}
	where $\He_n(E)$ is the Hermite polynomial in the probababilist's notation defined as:
	\begin{equation}
		\begin{aligned}
			\He_n(E) = (-1)^n \exp(E^2/2) \frac{d^n}{dE^n}  \exp(-E^2/2).
		\end{aligned}
	\end{equation}
	The expansion in the form in \cref{eq:def_gram_charlier} is used for a distribution function with zero mean and unit variance; any distribution can be cast in this form upon translating and rescaling. The coefficients of the Gram-Charlier expansion are given by
	\begin{equation}\label{eq:gram_charlier_coefs}
		c_n = \frac{(-1)^n}{n!} \int dE \; P(E) \; \He_n(E).
	\end{equation}
	The coefficients can be written in terms of the moments of the distribution 
	\begin{equation}
		\mu_n = \int dE \; P(E) E^n = \langle E^n \rangle,
	\end{equation}
	once the Hermite polynomials are expanded. A list of the coefficients is given in \cref{tab:series_coeffs}. 
	
	The Edgeworth series is obtained by regrouping the same terms from a Gram-Charlier:
	\begin{equation}
		\begin{aligned}
			\tilde{P}(E) = &\frac{\exp(-x^2/2)}{\sqrt{2\pi}} \bigg[  1+ \frac{\kappa_3}{6} \He_3(E) \\
			+ &\left(\frac{\kappa_4}{24} \He_4(E) + \frac{\kappa_3^2}{72} \He_6(E) \right) + \ldots  \bigg] 
		\end{aligned}
	\end{equation}
	where $\kappa_n$ is the $n$-th cumulant of the distribution.
	For the general prescription for obtaining the terms and also explicit forms for more terms, see \cref{app:edgeworth series}.
	
	\begin{table}[]
		\centering
		\begin{tabular}{|c|c|c|}
			\hline
			order  & Gram-Charlier coefficient \\
			\hline
			3 & $-\frac{1}{3!}\mu_3$ \\
			4 & $\frac{1}{4!}\left[\mu_4-3\right] $ \\
			5 & $\frac{1}{5!}\left[-\mu_5+10 \mu_3 \right]$ \\
			6 & $\frac{1}{6!}\left[\mu_6-15 \mu_4+30\right]$ \\
			7 & $\frac{1}{7!}\left[-\mu_7+21 \mu_5-105 \mu_3 \right]$ \\
			8 & $\frac{1}{8!}\left[\mu_8-28 \mu_6+210 \mu_4-315\right]$ \\
			\hline 
		\end{tabular}
		\caption{Coefficients defined in \cref{eq:gram_charlier_coefs}}
		\label{tab:series_coeffs}
	\end{table}
	
	In practice, the above series expansions may exhibit negative distribution values or artificial rapid oscillations. This happens mostly in cases where the energy distribution is far from Gaussian -- for example, when large gaps are present in the spectrum. Generally, the series will converge if the approximated distribution function falls faster than $\exp(-E^2/4)$~\cite{blinnikov1998expansions}. 
	
	The convergence condition is satisfied for a bounded energy spectrum, but the discreteness of the true energy distribution, from which the moments are calculated, can cause the series approximation to show rapid oscillations at higher orders. Moreover, we have seen in our numerics that Gram-Charlier series is generally more well-behaved than Edgeworth series.
	
	Therefore, to obtain a series approximation for the energy distribution, it suffices to calculate the moments $\langle E^n \rangle$. This can be done directly from knowledge of the classical wavefunction $\ket{\psi}$ and the system Hamiltonian $H$, although it can become computationally intensive. In this work, we mainly employ an MPS representation of the states and a matrix product operator (MPO) expression for the Hamiltonian to compute moments. We find that this does not impose a large computational overhead, as acting with the Hamiltonian MPO on the solution MPS even multiple times is not a prohibitive computational task.

	\subsubsection{The resolvent method}
	
	The energy distribution as defined in \cref{eq:def_energy_dist} can be thought of as the imaginary part of a particular Green's function, $P(E,\eta) = - \text{Im}\, G(E,\eta)$, with the Green's function defined as
	\begin{equation}\label{eq:green_resolvent_form}
		G(E,\eta) = \frac{1}{\pi}\bra{\psi} \frac{ 1 }{H-E + i\eta} \ket{\psi}.
	\end{equation}
	Transforming to the Lehmann representation by inserting the resolution of the identity, we find
	\begin{equation}
		P(E,\eta) = -\frac{1}{\pi} \sum_n \, |\langle E_n| \psi\rangle|^2 \ \text{Im} \left( \frac{ 1 }{E_n - E + i\eta} \right),
	\end{equation}
	and using the fact that 
	\begin{equation}
		\text{Im} \, \frac{1}{E_n-E+i\eta} =  \frac{- \eta}{ (E_n-E)^2 + \eta^2 },
	\end{equation}
	we can see that computing $- \text{Im}\, G(E,\eta)$ is equivalent to~\cref{eq:def_energy_dist} for a Lorentzian kernel with broadening $\eta$. Thus any method that can calculate $G(E, \eta)$ can be used to find the approximate energy distribution. We refer to calculating the energy distribution through its associated Green's function as the \textit{resolvent method}.

	We solve for the above Green's function through a DMRG-like variational method that was introduced in Ref.~\cite{jeckelmann2002dynamical} and then improved in Ref.~\cite{ronca2017time}. The method uses the MPS wavefunction form, and performs DMRG-like sweeps to evaluate the Green's function. This means that if we want to use the resolvent method to assess the quality of a candidate wavefunction, the state must be transformed into MPS form.
	
	The method works as follows: we define the state $\ket{\varphi}$ that satisfies:
	\begin{equation}
		\begin{aligned}
			&\ket{\varphi} = \frac{1}{\pi} \frac{1}{ (H-E+i\eta)}\ket{\psi},\\
			\Rightarrow \ \ & \pi (H-E+i\eta) \ket{\varphi} = \ket{\psi}.
		\end{aligned}
	\end{equation}
	The Green's function can be written as the overlap $G(E,\eta) = \langle \psi | \varphi \rangle$. 
	Now, defining $\ket{Y}$ as \cite{jeckelmann2002dynamical}:
	\begin{equation}
		\left[ (H-E)^2 + \eta^2 \right] \ket{Y} = -\frac{\eta}{\pi} \ket{\psi}.
	\end{equation}
	From this equation, we see that $\text{Im} \; G = \langle\psi|Y\rangle$.
	Finding the overlap of the above equation with $\ket{Y}$, a DMRG-like algorithm is used to minimize the resulting functional \cite{jeckelmann2002dynamical,ronca2017time}:
	\begin{equation}
		\bra{Y} \left[ (H-E)^2 + \eta^2 \right] \ket{Y} + \frac{\eta}{\pi} \; \langle Y \ket{\psi}.
	\end{equation}
	We rely on the package Block2 \cite{zhai2021low, Block2, zhai2023block2} to carry out the calculation of the Green's function in \cref{eq:green_resolvent_form}.
	Even though the method is implemented in the MPS language, it is general because we have methods for transforming other forms of states into MPS.
	
	\subsubsection{Coarse QPE}\label{sec:coarse_qpe_distribution}
	
	Finally, we describe how using low-precision QPE, which we refer to as \textit{coarse} QPE, can be used to build approximate energy distribution function; this is based on using the QPE kernel:
	\begin{equation}
		f(E) = \frac{1}{2^{2k}} \frac{\sin^2(\pi \, 2^k E)}{\sin^2(\pi E)},
	\end{equation}
	in \cref{eq:def_energy_dist}, for some integer $k$ that is the number of digits in the coarse QPE measurement.
	
	One possible way to obtain an energy distribution is through performing QPE multiple times and obtaining the statistics of the outcomes and then trying to reconstruct the energy distribution from that.
	As a refinement of this process and to diminish bias, for each QPE round, we add a random constant to the Hamiltonian. The constant $c$ is chosen to lie in the interval $[0,2^{-k})$. After the measurement is performed and an integer $x_m$ is observed, we take the measured energy to have the value $2^{-k} x_m - c$. This ensures that all real values can be sampled, making it possible to approximate the energy distribution using smoothing methods such as kernel density estimation~ \cite{wand1994kernel}. 
	As is well known, in kernel density estimation with a suitable choice of the broadening factor, the error scales as $1/M^{4/5}$, where $M$ is the number of samples (see \cref{app:kde} for more details).

	\begin{figure*}
		\centering
		\centering
		\includegraphics[height=0.27\textwidth]{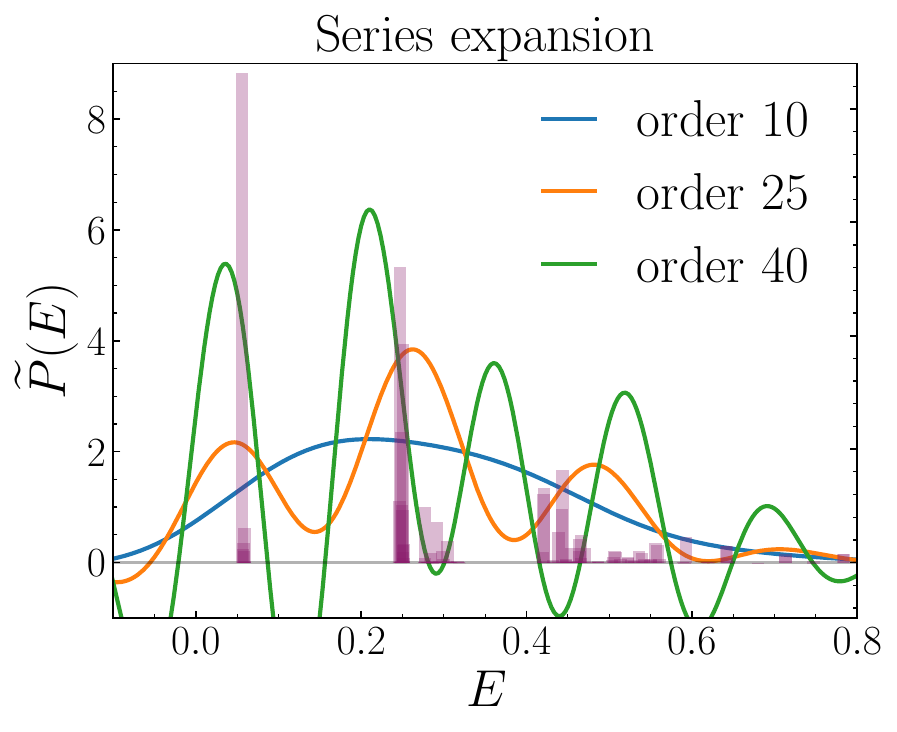}
		\includegraphics[height=0.27\textwidth]{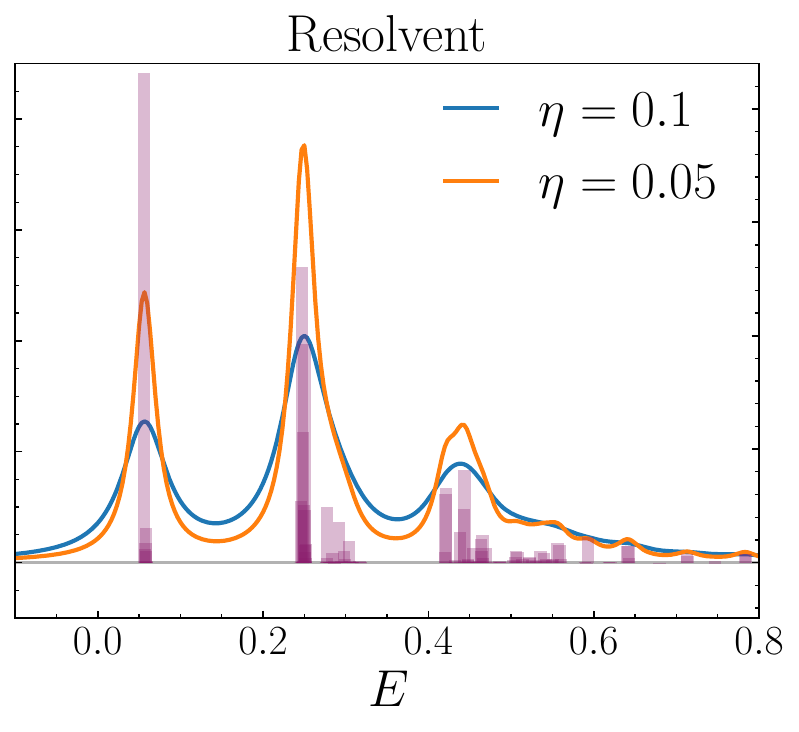}
		\includegraphics[height=0.27\textwidth]{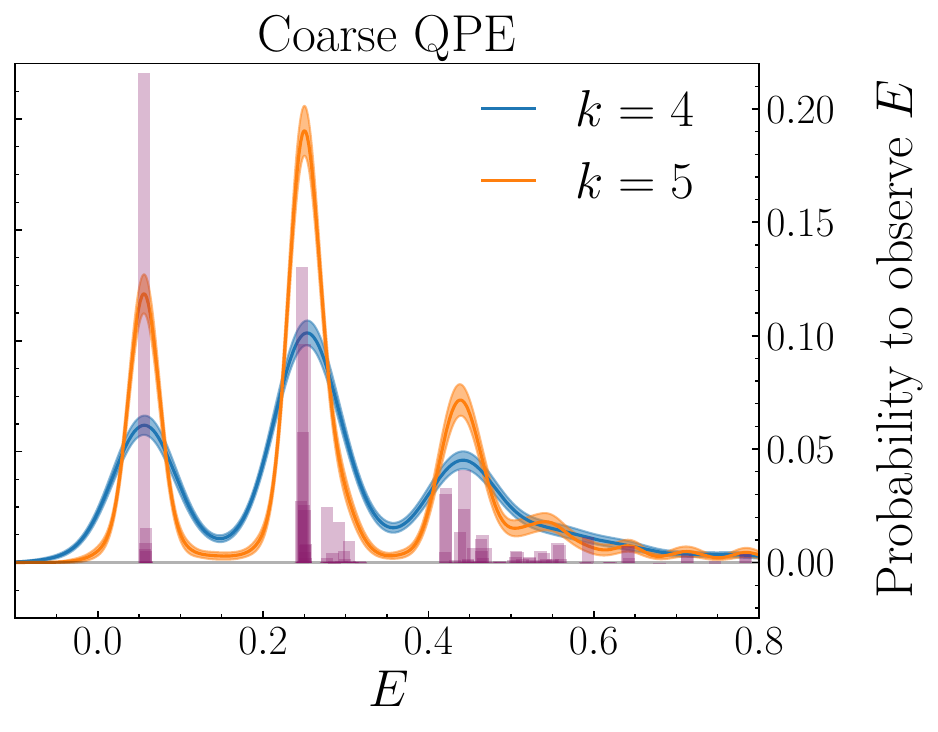}
		\caption{
			Energy distribution of an initial state which is the sum of three Slater determinants for the Hamiltonian of a hydrogen chain of length 6 within the STO-6G basis set. The bonds are extended to have the value of $5 a_0$, where $a_0$ is the Bohr radius.
			The exact overlap of the state with all of the eigenstates of the Hamiltonian is also calculated for reference.
			For each panel, two vertical axes are used: the left vertical axis corresponds to the energy distribution and the right one corresponds to the probability ($|\text{overlap}|^2$) of each Hamiltonian eigenstate in the expansion of the initial state.
			The horizontal axis energy is obtained from the actual energy in Hartree, by rescaling by a factor of $1/3$ and then translating by the value of $+1$, so that it lies between 0 and 1.
			Left: energy distributions obtained through the Gram-Charlier series. As the order is increased, more features are captured but also for higher orders unphysical oscillations also start to occur.
			Middle: energy distributions obtained through the resolvent method for two $\eta$ values. A bond dimension of 100 is used for the calculation of all the points of the two curves. For this particular system, the exact results with these $\eta$ values closely match these curves.
			Right: energy distributions obtained through the coarse QPE approach for 4 and 5 QPE qubits. A number 100 different realizations of the 50-measurement runs for both of the cases are considered and the mean and standard deviation are calculated. Kernel density estimation with a Gaussian kernel is also used for smoothing the curve with the broadening factor of $2^{-k}$. The resulting mean and error are shown as the solid curve and the shaded area around it, respectively. \label{fig:h6_dist_three_methods}
		}
	\end{figure*}
	
	\subsubsection{Numerical example}
	Here, we consider a Hydrogen chain with 6 Hydrogen atoms at a bond length of $5 a_0$, where $a_0$ is the Bohr radius (see \cref{sec:H_chains_numerics} for more details on this class of systems). We calculate the exact spectrum and also find the energy distribution of an initial state of our choice using the above method. The state that is used has a form as a sum of Slater determinant as the following: 
	\begin{equation}
		\begin{aligned}
			\ket{\psi_0} =& 0.86\ket{2,2,2,0,0,0}\\
			- & 0.36\left(\ket{\beta,2,\alpha,\alpha,0,\beta} + \ket{\alpha,2,\beta,\beta,0,\alpha}\right),    
		\end{aligned}
	\end{equation}
	in the basis of Hartree Fock orbitals. The coefficients are directly taken from the corresponding terms in the exact ground state but the state is normalized.
	
	The above three methods are used and energy distributions obtained are shown in \cref{fig:h6_dist_three_methods} (see the caption for the details of implementation of the methods). All three methods for this particular example show that some useful information can be obtained. But among the two classical methods we see that the resolvent method is more reliable, especially because the series expansions can show uncontrolled oscillations for larger orders.

	\subsection{Using the energy distribution for estimating lowest-energy outcomes}\label{sec:lowest_outcome}
	Assuming we have access to the approximate energy distribution of the initial state, we can use it to approximate the distribution of QPE outcomes. For a number of $k$ digits and an integer outcome $x_m$, it can be calculated as follows:
	\begin{equation}\label{eq:distribtuion_qpe_outcomes}
		\tilde{P}(x_m) = \int dE \; P(E) \; \frac{1}{2^{2k}} \left(\frac{\sin^2(\pi 2^k E)}{ \sin^2(\pi [E - x_m/2^k ])} \right).
	\end{equation}
	Note that this is a discrete distribution function. We define the variable $E_o = 2^{-k} x_m$ with the distribution function 
	$P(E_o) = \tilde{P}(x_m) /  2^{-k} $.
	In the limit of $k \to \infty$, this distribution approaches the underlying energy distribution and henceforth we denote this by $P(E)$ for notational simplicity.
	
	We study the statistics of the best energy achievable through repeated QPE measurements. Referring to $K$ QPE outcomes as $E^{(1)},\ldots,E^{(K)}$, we focus on the distribution of the smallest observed energy, $E_{\min, K} = \min(E^{(1)},\ldots,E^{(K)})$. It is straightforward to calculate the cumulative distribution function (CDF) of this variable at energy $E$ as:
	\begin{equation}\label{eq:min_N_cumul}
		C_{\min, N}(E) =  1-\left(1 - p_<(E)\right)^K.
	\end{equation}
	This is the probability that at least one outcome from $K$ rounds of QPE lies below $E_0$. Here we have defined 
	\begin{equation}
		p_<(E) = \int_{-\infty}^{E} dE' \; P(E'),
	\end{equation} 
	which is the probability of a single outcome lying below $0$. Upon differentiating the CDF with respect to $E$, we obtain the probability distribution function of $E_{\min, K}$, which reads:
	\begin{equation}
		P_{K}(E) = K P(E) \left(1 - p_<(E)\right)^{K-1}.
	\end{equation}
	One simple measure of state quality is the mean value of this distribution: $\int dE \; P_{K}(E) E$.

	\bigskip
	\section{Quantum refining}
	\label{sec:quantum-refining}
	After the classically optimized ansatz state is implemented on the quantum computer, there is possibility for further quality improvement by using a quantum algorithm to filter out some of the remaining high energy weights. This is beneficial only if a cheap quantum refining procedure is possible. We show that this is indeed the case in this section. Also, we show that another subtlety with QPE, namely the leakage problem, can be addressed through the quantum refining process.
	
	We focus on two main methods: coarse QPE \cite{berry2018improved} and quantum eigenvalue transformation of unitary matrices with real polynomials (QETU) \cite{dong2022ground}. The latter is chosen as a representative of the polynomial-based algorithms \cite{ge2019faster,lin2020near,dong2022ground}. 
	
	\subsection{Coarse QPE with postselection}
	Here we consider an implementation of QPE with less digits of precision than in the final scheme, which for example may be targeting chemical accuracy. In each coarse QPE measurement, if the outcome lies outside a set of predetermined low values, the state is discarded and the algorithm is restarted. The assessment of what QPE outcomes are considered small can be based on the energy distribution of the implemented state. 
	
	To see to what extent large energies are suppressed after postselecting on low-energy QPE outcomes, consider a setting in which we postselect an outcome $x_m$ when QPE with $k$ digits is performed on a state $\ket{\psi} = \sum_n c_n \ket{E_n}$. The probability for each component $E_n$ in the resulting state will be
	\begin{equation}
		P(n) = |c_n|^2 \, \frac{1}{2^{2k}} \left(\frac{\sin^2(\pi 2^k E_n)}{ \sin^2(\pi [E_n - x_m/2^k ])} \right).
	\end{equation}
	
	If the standard deviation of the initial state energy distribution is small compared to the span of the spectrum of the Hamiltonian, we can approximate the denominator in the above factor by a Taylor expansion. The weight after measurement then is suppressed as the inverse square distance from the measurement outcome: $\sim |2^k E - x_m|^{-2}$.
	
	This shows that if the precision of the coarse QPE and the postselection values are chosen appropriately, the high energy weight of the distribution can be well suppressed. 
	
	\subsection{QETU filtering}
	
	Using QETU~\cite{dong2022ground} or other polynomial-based methods~\cite{lin2020near,ge2019faster}) aim to implement a function of the Hamiltonian that retains low energies and filters high energies. In short, the method consists of a quantum signal processing circuit~\cite{low2017optimal,martyn2021grand} that implements a unitary matrix that block encodes a function $f(H) = P(\cos(H/2))$, where $H$ is the Hamiltonian of interest and $P$ is an even polynomial of degree $d_P$. We need $f(H)$ to be designed in a way so that low energies are retained and high energies are filtered, for instance using an approximate step function.
	Details of how this can be done are discussed in the appendix.
	
	The cost of implementation is directly given by the degree of $P$; more precisely, the number of times that one queries the unitary $U=e^{-iH}$ is exactly $d_P$. In order for the filtering to be successful, this degree should have a scaling $O(\Gamma^{-1} \log\epsilon^{-1})$, where the error $\epsilon$ in the polynomial approximation is small enough, and $\Gamma$ is the energy scale over which the transition in the function $f(H)$ needs to occur.
	Apart form the above asymptotic scaling, in practice, we choose the degree by examining how good of a filtering function is achieved.
	
	\subsection{Cost of implementing quantum refining methods}\label{sec:refining_costs}
	A simple analysis of the asymptotic cost of the above methods can be done by the following consideration: to suppress a high degree of weight at unwanted high energies, and to keep low-energy weights mostly intact, we need to differentiate energies separated by values of the order of the standard deviation $\sigma_E$ of the energy distribution of the initial state. 
	
	This means that in a coarse QPE setting, we need the resolution $2^{-k}$ to be of the order of $\sigma_E$; on the other hand, in a QETU setting, again we need a function $f(H)$ that discerns energies of the order of $\sigma_E$, i.e.,~$\Gamma \sim \sigma_E$. Thus, both of the methods require a number of queries to $e^{-iH}$ (or related unitary) that scales as $O(1/\sigma_E)$. We expect the target precision $\epsilon$ to be considerably smaller than $\sigma_E$, so the cost of filtering should in principle be much lower than for the final energy estimation algorithm, which would have a scaling $O(1/\epsilon)$. This shows that quantum refining for further improvement of the state quality is viable through these methods.
	
\begin{figure}[t]
		\centering
		\includegraphics[height=0.28\textwidth]{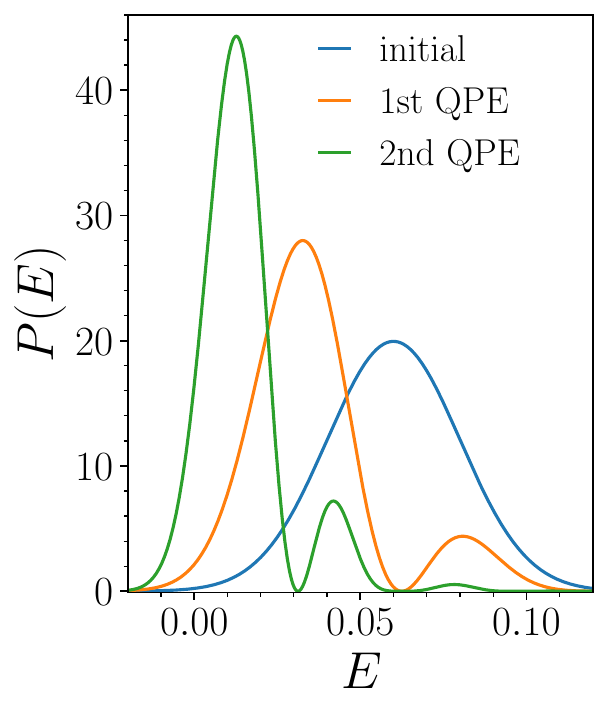}
		\includegraphics[height=0.28\textwidth]{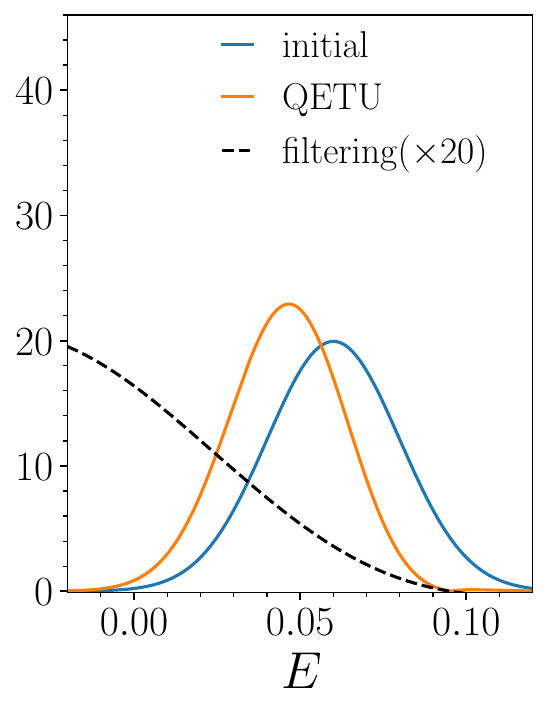}
		\caption{ An initial state with a Gaussian energy distribution is subject to quantum refining. The energy is expressed in some dimensionless units.
			Left: two rounds of coarse QPE are performed and the resulting energy distributions are plotted.
			Right: QETU with a step-like filtering function is performed and the resulting distribution along with the filtering function are depicted.
			More details of this procedure is presented in \cref{app:qetu_vs_cqpe}.}
		\label{fig:filtering}
	\end{figure}

	While the asymptotic costs are the same, by examining simple concrete examples, we find that coarse QPE works appreciably more efficiently. In particular, we consider the case of a Gaussian energy distribution for the initial state, and study the effect of quantum refining via coarse QPE and QETU on it; even though this can be an artificial construction, we can capture the essential factors of comparing polynomial based algorithms with coarse QPE in it. Plots of energy distribution after quantum refining are presented in \cref{fig:filtering}, with coarse QPE on the left panel and QETU on the right. We can see that the ultimate state obtained through coarse QPE has a much higher quality when compared with the one obtained through QETU.  
	More details on the processes are presented in \cref{app:qetu_vs_cqpe}. There, we show that this performance of coarse QPE is achieved even with a lower cost compared with QETU.
	
	Given this example and similar other constructions, we come to the conclusion that QPE has generally a lower cost in practice. This is mostly due to the fact that constructing polynomials with sharp jumps for the QETU algorithm requires high degrees, resulting in high costs. Thus we pick coarse QPE as our method of choice for performing the quantum refining stage of the algorithm.

	\subsection{Case study: The QPE leakage problem}\label{sec:leakage_intro}
	Here, we consider a known problem that is usually discussed in the literature as changing the cost scaling of QPE to a $1/p_0^2$ behavior instead of $1/p_0$, where $p_0$ is the square of the overlap of the initial state with the ground state (see e.g.~section I.A of \cite{lin2022heisenberg} and Appendix A of \cite{ge2019faster}).
	This scaling can occur when long tails of QPE kernels placed at higher energies contribute outcomes at low energies, potentially even below the ground state value.
	It is argued in Ref.~\cite{ge2019faster} that in order to prevent this from happening, longer evolution times that scale as $\frac{1}{p_0}$ in each QPE round should be used. Since a total of $\frac{1}{p_0}$ rounds are required to obtain precisely the ground state, the overall cost in this argument scales as $\frac1{p_0^2}$.
	We analyze this problem, which we call the QPE \textit{leakage problem}, based on our energy distribution approach and discuss how it can be diagnosed. 
	We also show that the problem can be circumvented using quantum refining of the initial state as discussed above, without a need to resort to large evolution time. 
	
	We first consider the problem in the conventional setting; starting with an initial state $\ket{\psi} = \sum_n c_n \ket{E_n}$, we would like to perform QPE and estimate the ground state energy $E_0$ with a tolerated  error $\epsilon$. The question is how many phase digits are required for this task. One requirement is to ensure that leakage is absent, meaning we need to estimate the probability of contributing an outcome below $E_0-\epsilon$ from all of the energy levels except the ground state:
	\begin{equation}\label{eq:total_probability_leakage_sum}
		p_{\text{leak}} = \sum_{n \neq 0} \sum_{x_j<x_{\text{upper}}} \frac{1}{2^{2k}} \left(\frac{\sin^2 (\pi \delta_n)}{ \sin^2 \left(\frac{\pi}{2^k} [x_n+\delta_n - x_j]\right)} \right),
	\end{equation}
	where $x_{\text{upper}}=\lceil 2^k (E_0-\epsilon) \rceil$, $2^k E_n = x_n + \delta_n$ with $x_n$ an integer satisfying $0 \leq x_n < 2^k$, and $0 \leq \delta_n < 1$. 
	For the leakage to be improbable, $p_{\text{leak}}$ should be small enough when compared with probability of the ground state in the initial state composition $p_0=|c_0|^2$. 
	Note that for the ground state itself to not leak beyond the threshold one can add an $O(1)$ number of more qubits to the phase register and discard their outcome \cite{nielsen2010quantum}.
	
	Up to an additive error of $O\left(\max \left[2^{-2k},(x_n-x_{\text{upper}})^{-2} \right] \right)$ , the single level leakage probability in \cref{eq:total_probability_leakage_sum} can be written as:
	\begin{equation}\label{eq:single_level_approx}
		p_{\text{leak}} (E_n) = \frac{\sin^2(\pi \delta_n)}{\pi^2} \frac{1}{x_n-x_{\text{upper}}+\delta_n}.
	\end{equation}
	See \cref{app:leakage} for more details.
	Having access to the energy distribution of the initial state, the total leakage probability in \cref{eq:total_probability_leakage_sum} can be approximated as:
	\begin{equation}\label{eq:prob_int_below_xupper}
		p_{\text{leak}} = \frac{1}{\pi^2 2^k } \int_{E_0+\epsilon} dE \; P(E) \  \frac{\sin^2(\pi 2^k E)}{E - \frac{ x_{\text{upper}} }{ 2^k}   } .
	\end{equation}
	
	A simple criterion for leakage, based on the above integral, can be derived for multimodal distributions; we focus on a unimodal distribution and the multimodal case is similar. Assuming an $O(1)$ probability is concentrated close to the main peak of the distribution, and that this peak is located at $E_p$, from the above integral the probability of leakage beyond the ground state can be approximated as $\frac{1}{2\pi^2 2^k} \frac{1}{E_p-E_0}$. We need this probability to be smaller than the probability of the ground state; this means that the number of phase digits should be chosen large enough so that $2^k = O \left( \left[p_0 \left( E_p - E_0 \right)\right]^{-1} \right)$. In general, we need to take the tolerated error $\epsilon$ in QPE also into account for $k$, and as a result we have: 
	\begin{equation}\label{eq:time_limit_approximation}
		2^k = O \left( \max \left( \left[p_0 \left( E_p - E_0 \right)\right]^{-1}  , \epsilon^{-1}
		\right) \right).
	\end{equation}

	Apart from the above setting, where higher energy states can contribute to QPE outcomes below the ground-state energy value, there is also a possibility for leakage when we are not aiming to necessarily obtain the ground state energy but striving to obtain better energy estimates using QPE. In such a setting, if a small QPE outcome is obtained in an energy region where there is actually not an appreciable weight, it is more probable for the outcome to be invalid as it likely happened due to leakage from higher energies. Such outcome should not be accepted as an estimate of the energy since it can in general be smaller than all the eigenvalues of $H$ with which the initial state has nonnegligible overlap; it can actually be below the ground-state energy of the system resulting in incorrect estimates.

	To quantify such a possibility, we first note that the distribution of the lowest outcomes of QPE was considered in \cref{eq:distribtuion_qpe_outcomes} through the use of the energy CDF; here, we use the CDF of QPE outcomes to study the possibility of leakage too. In particular, at an energy of interest, we can compare the CDFs of energy and QPE outcomes (with the desired number of digits $k$). If the QPE outcomes CDF is considerably larger than the energy CDF, this signifies a high probability of leakage contamination of results around or below that energy value. This is especially important for the region in which the energy CDF is of the order $1/N$ for a QPE measurement with $N$ repetitions as this is where the smallest outcome is expected to appear. An ultra-precise QPE will result in an energy close to this region but if the QPE outcomes CDF is large, {\it lower precision} QPE can contribute smaller outcomes, and those can only be due to unwanted leakage from higher energies and thus should be avoided.
	
	Given these two treatments of the possibility of leakage, we see that knowledge of the energy distribution function enables us to identify situations in which QPE leakage is not insignificant. If leakage is present, in principle in both of the cases, with choosing QPE precision high enough, the leakage probability is managed; however this induces extra cost for each QPE round, as with increasing the number of QPE digits $k$, the cost rises exponentially $2^k$. In the following we show that quantum refining of the initial state can be used to manage the leakage possibility without a need to use higher precision.

	\subsubsection{Mitigating the leakage probability}
	The above analysis shows that if the spectral weight in some region, that is responsible for the leakage, is suppressed through some means, the leakage probability could also be suppressed with it. We have seen in \cref{sec:refining_costs} that for a small cost compared with that of the most precise QPE measurement, high energy weights can be filtered with quantum refining. Thus, QPE leakage can in principle be mitigated through quantum refining, but it is a matter to be studied on a case-by-case basis based on the details of the energy distribution at hand. This will be very helpful as it removes the need to perform time evolution for times of the order $p_0^{-1}$ for the ultimate QPE measurement.

	As an example, consider an energy distribution having a multimodal structure, one can identify the peaks -- accumulating $O(1)$ spectral weight in their vicinities -- which are responsible for leakage through analyses discussed above; one can then perform a quantum refining in the form of coarse QPE, so that those peaks are close to discarded outcomes, and thus will lose a substantial weight after the process. This can substantially decrease the leakage probability. Such procedure is illustrated in a concrete example in \cref{app:qetu_vs_cqpe} (last paragraph).

	Note that this means the quantum refining step of our state preparation algorithm is capable of reducing the cost of the whole algorithm, not only by lowering the number of required repetitions of the most precise QPE measurement, but also by decreasing the cost of each single round by mitigating the leakage problem when it is present.

	\section{Numerical demonstrations}
	\label{sec:numerics}
	In this section, we showcase our complete initial state preparation algorithm for a variety of molecules. Through the numerical examples, we explore how viable it is to prepare good-enough initial states for complex molecules to be studied with a quantum algorithm. Of particular interest are situations where, on the one hand, classical methods struggle to give a good energy estimate, but at the same time one can still prepare a good-enough initial state for quantum energy estimation. We call such problems Goldilocks problems, a concept formalized through the energy distribution.
	
	\subsection{Goldilocks molecules}
	\label{sec:goldilocks}
	
	As we argued in \cref{sec:energy-distribution}, the energy distribution picture can be used to estimate whether a quantum algorithm can improve a given initial state's classical energy estimate. Using this concept, we can categorize energy estimation {\it problems} based on the hardness of preparing a good enough initial state for performing quantum energy estimation routines (e.g.~QPE). 
	Such a classification is of course tied to the available budget for initial state preparation: here we take the budget to be unspecified for the sake of generality.
	With a given budget for state preparation,
	one can have easy, intermediate, and hard problems: an easy problem is one in which a very high quality initial state with large accumulated weight over low energy parts of the spectrum of the Hamiltonian is possible to prepare on a quantum computer; on the other hand, a hard problem will be one in which with a given budget, it is only possible to prepare a poor quality state with negligible weight over low energies of the Hamiltonian. Note that these depend on the given budget for state preparation. In-between, there ought to be problems of intermediate hardness, where there is some non-negligible, but also not too large weight over low energy parts of the spectrum. 
	
	We argue that in all likelihood, it is only \textit{possible} to obtain quantum advantage in quantum ground energy estimation over classical computational methods in \textit{intermediate} problems. This is because for hard problems, by definition one cannot perform quantum routines effectively; and for easy problems, it is very likely to find a good classical energy estimate which is highly challenging to beat using quantum algorithms. The question of whether there is quantum advantage in an intermediate problem remains open: the energy distribution allows this question to be explored computationally. 
	
	To make this more concrete, on an energy axis, let us mark the best classically achieved energy by $E_T$. Assuming we have access to the energy distribution of our initial state, we can calculate its accumulated weight for energies below $E_T$ and call it $p_{_{<E_T}}$. If $p_{_{<E_T}}$ is large enough -- that is, large enough so that performing an accurate QPE measurement a number $O(p_{_{<E_T}}^{-1})$ times is within the available QPE budget -- there is possibility for quantum advantage. We call such situation a \textit{Goldilocks problem} -- one in which there is possibility for improving the classical ground energy estimate using QPE. Given our numerical results for the Cr$_2$ and Fe$_4$S$_4$ molecules in the following sections, we expect that complexes including several transition metal centers, if studied within in an appropriate active space, could indeed present such Goldilocks problems.
	
	In the remainder of this section, we will study several different quantum chemical systems numerically and exhibit the ideas and discussions in the previous sections concretely in those systems. For all calculations in this section, we used the publicly available software library Overlapper \cite{overlapper}.
	
	\begin{figure*}[!t]
		\centering
		\includegraphics[width=0.32\textwidth]{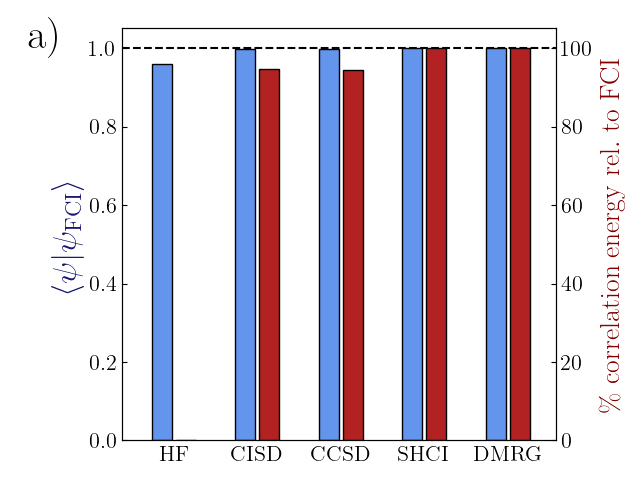}
		\includegraphics[width=0.32\textwidth]{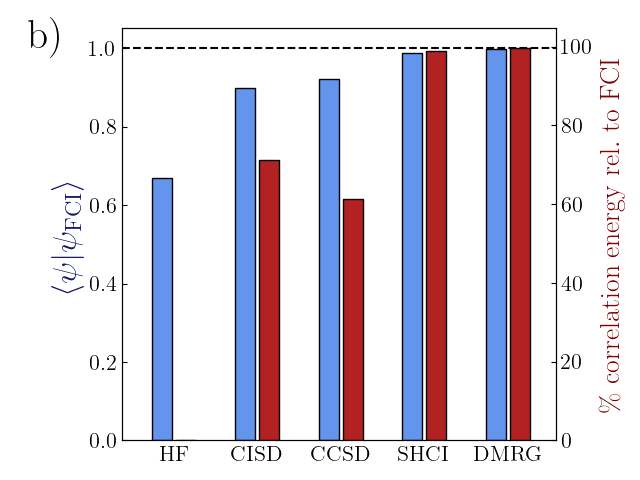}
		\includegraphics[width=0.32\textwidth]{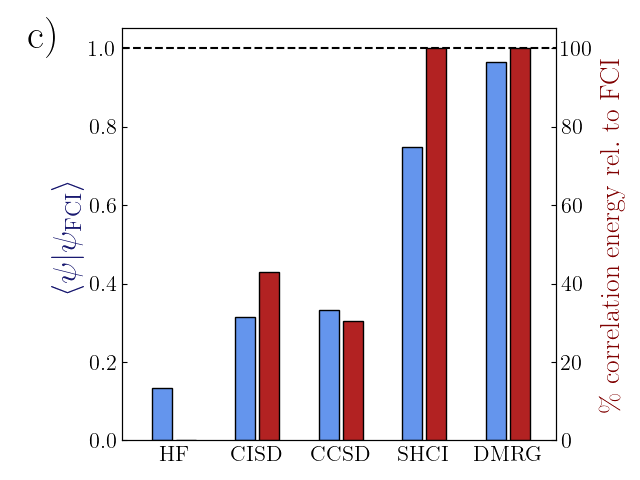}
		\caption{Percent recovered correlation energy relative to FCI (red) and overlaps $\langle \psi|\psi_{\text{FCI}}\rangle$ (blue) for H$_{12}$ chain in the STO-6G basis with varying bond length, with overlaps computed relative to the FCI solution: (a) bond length near equilibrium of H$_2$; b) twice the bond length of H$_2$; c) four times the bond length of H$_2$. CASCI / MRPT are excluded because they give the exact solution.}
		\label{fig:hchain-ovlps}
	\end{figure*}
	
	\subsection{Hydrogen chains}\label{sec:H_chains_numerics}
	
	We begin by studying the hydrogen chain model system in the minimal STO-6G basis (\cref{fig:hchain-ovlps}), varying the bond length to increase correlations and evaluating the overlaps and energies of different methods relative to the exact solution from full configuration interaction (FCI). All methods including DMRG are executed in the $S_z$ symmetry mode, i.e., conserving the spin projection on the $z$-axis. Here and in later figures, the overlap $\langle \psi | \psi_{\text{FCI}} \rangle$ is computed by first bringing the output of all methods to the SOS form: in all cases (especially DMRG), we make sure that the SOS form of the wavefunction includes enough determinants to capture above 99\% of the weight of the original state. Near equilibrium, as in panel (a), all methods perform equally well: but as the bonds get stretched and static correlation increases (panels (b)-(c)), SHCI and DMRG clearly emerge as leaders in terms of overlap, while the Hartree-Fock state performs poorly. While in practice CCSD is able to recover a substantial portion of the dynamic correlation energy of the system, its energy estimates are not directly linked to the quality of the wavefunction because the energy CCSD computes is non-variational. To place the CCSD energy estimates on equal footing with the other methods, we instead plot the variational energy of the associated CCSD ansatz that we aim to implement on a quantum computer: this amounts to expanding the CCSD wavefunction as a SOS and truncating it to second order. We find that such a truncated CCSD wavefunction is only marginally improved compared to CISD. This is expected: since CCSD's internal optimization routine is geared towards minimizing a non-variational energy, this can easily lead towards inaccurate wavefunctions. 
	
\begin{figure}[!t]
		\includegraphics[width=0.47\textwidth]{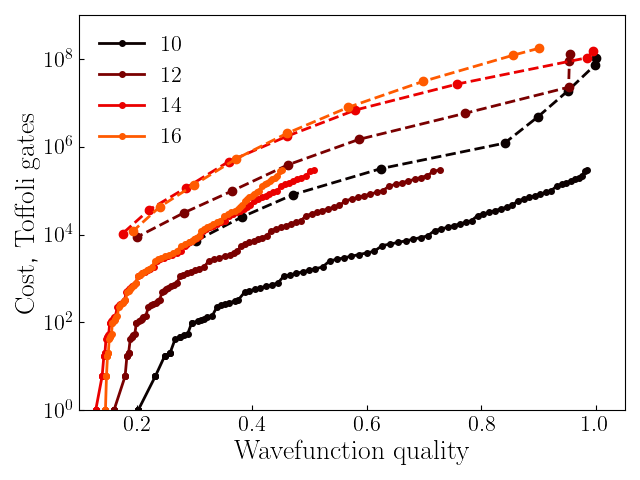}
		\caption{Comparing the cost of implementation, in terms of Toffoli gates, of the DMRG solution in SOS form (solid lines) and MPS form (dashed lines), for the hydrogen chain of different lengths, as a function of the prepared wavefunction overlap with the ground state. The bond length has been stretched to four times the equilibrium length. As the system gets larger, the MPS form continues to allow the preparation of wavefunctions with high quality. The finite steps in the SOS cost curve are due to rounding of $\log(D)$ factors to the nearest integer.}
		\label{fig:hchain-sos-vs-mps}
	\end{figure}

	Hydrogen chains are also convenient for comparing the SOS and MPS forms of the initial state in terms of their quality versus implementation Toffoli cost curves. For this comparison, we obtain the system ground state with DMRG, then process it in two ways: for the SOS form, we do reconstruction and then repeatedly truncate the number of Slater determinants; for the MPS form, we compress the MPS to smaller and smaller bond dimensions while continuously computing the overlap to the original state. For implementation cost, we use the expressions derived in \cref{sec:quantum-implement}: the cost is mainly set by the number of determinants $D$ for SOS and by the bond dimension $\chi$ for MPS. The results for hydrogen chains of varying sizes are shown in \cref{fig:hchain-sos-vs-mps}. 
	
	The MPS form appears to be more expensive relative to the SOS form for the cases shown even though the system is one dimensional, which means that MPSs should perform very well in representing the ground state. This signifies the efficiency of the SOS method developed in this work. Notice that DMRG can still be used as the method of choice for the classical ground-state search in such cases, however, it might beneficial to transform the result into an SOS form and then implement on a quantum computer.
	Furthermore, we should also note that the SOS cost is seen to increase exponentially with wavefunction quality, especially in larger chains, as more and more determinants are needed to accurately represent the ground state. In fact, the reason why the SOS curve does not reach perfect wavefunction overlap for larger chains is because of the extreme memory requirements for storing all the determinants arising in reconstruction beyond a certain cutoff. Thus there could be advantages to implementing the state in the MPS form in large, strongly-correlated systems. This needs to be explored more in future works.
	
\begin{figure}[!t]
		\centering
		\includegraphics[width=0.47\textwidth]{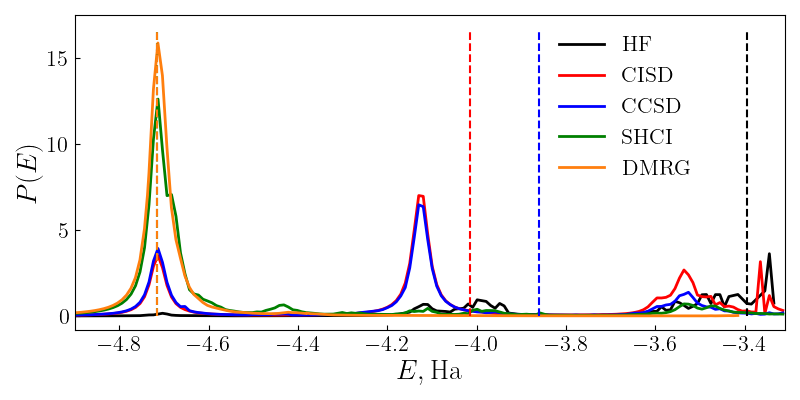}
		\caption{Energy distributions for the initial states from the methods studied in \cref{fig:hchain-ovlps}, for the hydrogen chain with 10 atoms with bonds stretched to four times the equilibrium length of the H$_2$ molecule. For CCSD, the energy reported is that of the truncated SOS state rather than the non-variational energy evaluated by the CCSD method. The calculations are done with the resolvent method with $\chi_{\text{calc}} = 200$ and $\eta = 0.02$. Here and in all later energy distribution plots, vertical dashed lines give the classical energy of the state of the same color. The DMRG and SHCI energies coincide.}
		\label{fig:hchain-edists}
	\end{figure}

	We next study energy distributions for all of the states studied in \cref{fig:hchain-ovlps}. We compute them using the resolvent method: for all non-MPS-based states, we first convert them to MPS form using subroutines from Overlapper \cite{overlapper}. Near equilibrium, all energy distributions are sharply peaked around the ground state (not shown). However, when bonds are stretched, the DMRG and SHCI states have significantly more weight in the lower-energy portion of the spectrum, as seen in \cref{fig:hchain-edists}. The classical energies of the states are shown with vertical dashed lines of the same colour.  Given that we are implementing the SOS forms of the states, we report the associated SOS state energy, not the energy reported by the classical computational method. This applies most importantly to CCSD: because of this consideration, the CCSD energy shown is actually higher than the CISD energy. While the CCSD state redistributes some of the weight towards lower-energy states relative to CISD, this is marginal. At the same time, the Hartree-Fock state has very little weight in the low-energy parts of the spectrum, including the ground state peak. Overall, we see that even though these energy distributions are approximate, by direct visual inspection the high-quality states can be identified -- without any need for a reference state, as with the overlap metric. Beyond this, we also get a much richer picture of how the weight is distributed across the energy range.
	
	\subsection{The N$_2$ molecule}
	
	Next we turn to molecules, starting directly with a system with an intermediate degree of correlations --- the N$_2$ molecule with stretched bonds ($r = 2.25 r_{N_2}$) in the cc-pVDZ basis with the effective core potential ccECP \cite{bennett2017new}, which is used to reduce the number of electrons in the problem. Since this system of 26 orbitals and 10 electrons is now beyond our capability for FCI, we use a highly-converged DMRG wavefunction ($\chi = 1000$) as the reference state, because it returns the lowest energy for this system (a few mHa below the SHCI solution). The CASCI and MRPT methods are carried out in an active space of CAS(10e, 12o). 
	
\begin{figure}[!t]
		\centering
		\includegraphics[width=0.47\textwidth]{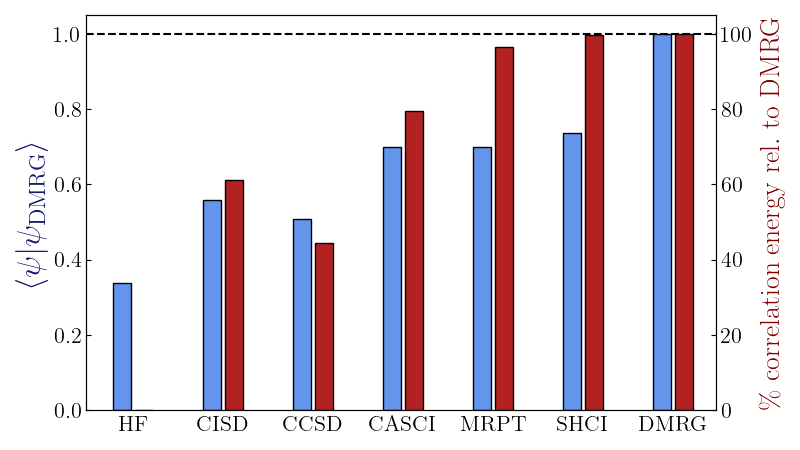}
		\caption{Percent recovered correlation energy and overlaps for the N$_2$ molecule in the cc-pVDZ basis, with its bond stretched to 2.25 the length, relative to the DMRG solution which obtained the lowest energy. The CASCI / MRPT methods are performed with the active space CAS(12o,10e), chosen directly from molecular orbitals around the Fermi level.}
		\label{fig:n2-2.25-ovlps}
	\end{figure}

	The energy-overlap bar chart in \cref{fig:n2-2.25-ovlps} shows that most methods, while they struggle to generate an excellent wavefunction, still provide reasonably good overlap to the reference state. Notice that while MRPT significantly improves the energy estimate, it does not improve the CASCI wavefunction quality. Another curiosity is that the SHCI wavefunction, while recovering nearly 100\% of the correlation energy, has an appreciably reduced overlap with the DMRG reference state. All of these observations reinforce the idea that energy is not a very reliable proxy for wavefunction quality.
	
\begin{figure}[!t]
		\centering
		\includegraphics[width=0.47\textwidth]{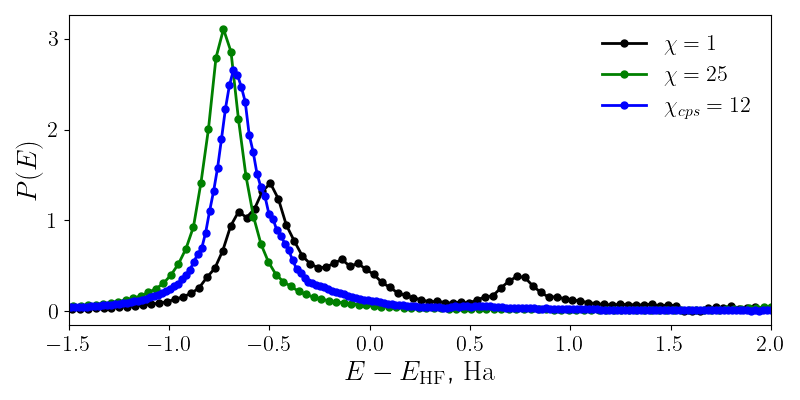}
		\caption{Energy distributions for DMRG-generated states for the N$_2$ molecule, with the bond length stretched to 2.25 times the equilibrium bond length. The calculations are done with the resolvent method with the reduced parameters $\chi_{\text{calc}} = 50$ and $\eta = 0.1$ to speed up the calculations.}
		\label{fig:n2-2.25-edists}
	\end{figure}

	For the energy distributions for this molecule, we choose to compare the states obtained from DMRG (see \cref{fig:n2-2.25-edists}). Namely, we compare the Hartree-Fock state ($\chi = 1$); an intermediate-quality state with bond dimensions $\chi = 25$ obtained through an increasing bond dimension schedule without compression; and finally $\chi_{cps} = 12$, a state which was obtained initially at $\chi = 250$ and then subsequently compressed to $\chi_{cps} = 12$ before the energy distribution was computed. Our best classical result is obtained with $\chi = 1000$. Once again, at a glance we notice that the Hartree-Fock state is of significantly worse quality than the other two states. A further interesting observation is that going to large bond dimensions and then compressing, as in the $\chi_{cps} = 12$ state, gives results nearly as good as the $\chi = 25$ state but with lower implementation cost. Beyond the details, this example shows that the energy distribution method allows us to characterize and compare the quality of initial states for physical molecules.
	
	\subsection{The Cr$_2$ dimer}
	
	We next turn to the Cr$_2$ molecule, which is an example of a strongly multireference system. To increase correlations even further, we stretch the dimer bond length by a factor of 1.8 relative to equilibrium. To study such a many-electron molecule with a limited computational budget, we focus on active spaces built around a limited set of orbitals, the $3d$-orbitals in Cr$_2$. We employ the atomic valence active space (AVAS) approach~\cite{sayfutyarova2017automated}, where molecular orbitals with the largest $d$-orbital overlap above a given threshold are selected for the active space.
	
\begin{figure}[!t]
		\centering
		\includegraphics[width=0.4\textwidth]{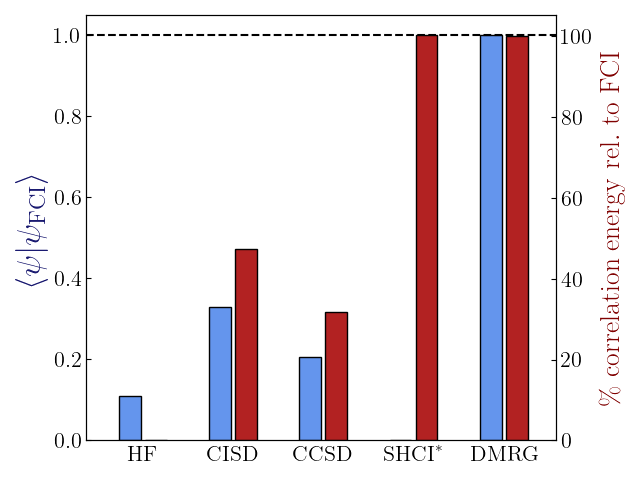}
		\caption{Percent recovered correlation energy and overlaps for the Cr$_2$ molecule with bond length 1.8 times the equilibrium length, in the cc-pVDZ basis in an active space of only $d$-type orbitals, relative to the FCI solution. Note that the SHCI solution could not be made to converge to a state with $S^2 = S_z = 0$: instead, it finds a state with non-zero total spin, which has an energy lower than the $S^2 = S_z = 0$ state, but also a vanishing overlap with the chosen FCI reference.}
		\label{fig:cr2-ovlps}
	\end{figure}

	The active space for Cr$_2$ focused on $3d$ orbitals can be as small as 10 active electrons in 10 active orbitals, written CAS(10e,10o). In an active space this small, a reference FCI solution can be obtained. As seen in \cref{fig:cr2-ovlps}, once again the DMRG method recovers the entirety of the correlation energy and, more importantly, produces a wavefunction with perfect ground-state overlap, whereas both CISD and CCSD clearly struggle in this multireference situation. Note that the DMRG calculations are now being carried out in $\text{SU}(2)$ mode, conserving total spin $S$, instead of only the projection $S_z$. As before, CCSD appears to recover most of the correlation energy whilst having a worse overlap than the CISD solution.  
	Notice that the SHCI solution gives nearly the correct energy but an incorrect wavefunction: in our calculations, we found it particularly challenging to obtain an SHCI solution with the correct total spin, i.e.~total spin of zero, the chosen FCI reference solution -- hence the vanishing overlap. 
	
	For the energy distributions, we again focus on the DMRG exact solution and its compressed versions with lower bond dimensions. The energy distribution gives an excellent account of the quality of the state during compression: not only are the different quality states easily distinguishable, it is also clear that even the poor-quality, highly compressed $\chi_{cps} = 4$ state still has a wide range of energies covered in its distribution, showing weight well below its mean (classical) energy.
	These energy distributions are an illustration of the Goldilocks state concept described in more detail in \cref{sec:goldilocks}: while the $\chi = 4$ state is clearly too poor to do QPE with, and the $\chi = 250$ state already has a good classical energy, the intermediate quality states with $\chi = 50-100$ would allow QPE to improve on their associated best classical estimate with only a small number of iterations. Notice also that while the quality of the $\chi = 100$ state is greater than that of the $\chi = 50$ state, their energies are nearly equal: this shows that it is possible to find initial states with improved quality without this corresponding to an improvement in the classical energy estimate.
	
	\begin{figure}[!t]
		\centering
		\includegraphics[width=0.47\textwidth]{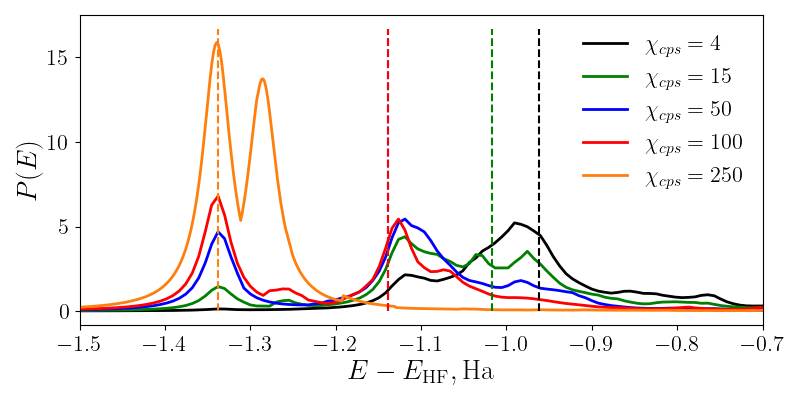}
		\caption{Energy distributions for different states of compression of the DMRG-generated ground state: starting with a well-converged state with $\chi = 500$, the state was slowly compressed to $\chi_{cps} = 250$ (orange) and then further to $\chi_{cps} = 100, 50, 15, 4$ (red, blue, green, and black respectively). As before, the calculator bond dimension used is $\chi_{\text{calc}} = 200$, with broadening $\eta = 0.02$. While this is a relatively large calculator bond dimension, the energy distribution seen here is still approximate and not fully converged with $\chi_{calc}$. However, even an approximate distribution already allows the evaluation of state quality, whilst having the advantage of being computable with reasonable resources. Compression, while reducing resources necessary for preparation, partially trades that off against the quality of the initial state.}
		\label{fig:cr2-edists}
	\end{figure}
	
	\subsection{[Fe$_4$S$_4$] core}
	
	Finally, we consider the situation in the Fe$_4$S$_4$ molecule core -- the 8-atom center extracted from the associated molecular iron-sulfur complex. The active space for this system is focused on the Fe $3d$ and S $3p$ orbitals: we follow the procedure outlined in Refs.~\cite{sharma2014low, lee2023evaluating}. Instead of Pipek-Mezey, we use the Cholesky method to split-localize the $\alpha$ molecular orbitals from a high-spin ($S_z = 5/2$ per Fe atom) restricted open-shell Hartree-Fock calculation in the cc-pVDZ basis, then select orbitals with Fe $3d$ and S $3p$ character by visual inspection.
	
	For this system, we consider four different states obtained with DMRG: three states obtained by converging DMRG at bond dimensions $\chi = 20, 50, 100$ respectively, and another obtained by converging a calculation at a high bond dimension of $\chi = 1000$ and then compressing that wavefunction down to $\chi_{cps} = 7$. While the highly compressed state is the most high-quality of the four, it has the worst classical energy estimate -- higher-energy admixtures balance out the weight at the lower energies. On the other hand, while the $\chi = 50$ state has a significantly improved energy relative to $\chi = 20$, its energy distribution is mostly unchanged, and its quality not significantly improved. Finally, the states $\chi_{cps} = 7$ and $\chi = 20$ have nearly the same energies, but the former has much more weight in the low energy part of the spectrum, making it much higher-quality. At the same time, all these states for the Fe$_4$S$_4$ core are built up from tens or hundreds of Slater determinants, with the coefficients of the largest contributing determinants being on the order of $10^{-2}$. This strongly suggests a single product state would be an exceedingly poor initial state in this situation. More specifically, the $\chi_{cps} = 7$ state has an overlap of $0.69$ with the ground state (obtained at $\chi = 1000$), while the largest contributing determinant in the $\chi = 1000$ state has weight $0.04$. Then in terms of probabilities to project on the low-energy $\chi = 1000$ state, the $\chi_{cps} = 7$ state gives $0.47$ while the best single product state gives $0.002$. Thus at least 300 times fewer iterations of QPE will be needed to project on this state with the only marginally more complicated initial state -- which translates into direct cost and runtime savings of more than two orders of magnitude. At the same time, the compression to $\chi_{cps} = 7$ ensures that the cost of implementing the improved initial state continues to be negligible compared to the main costs of energy estimation \cite{delgado2022simulating}.
	
\begin{figure}[!t]
		\centering
		\includegraphics[width=0.47\textwidth]{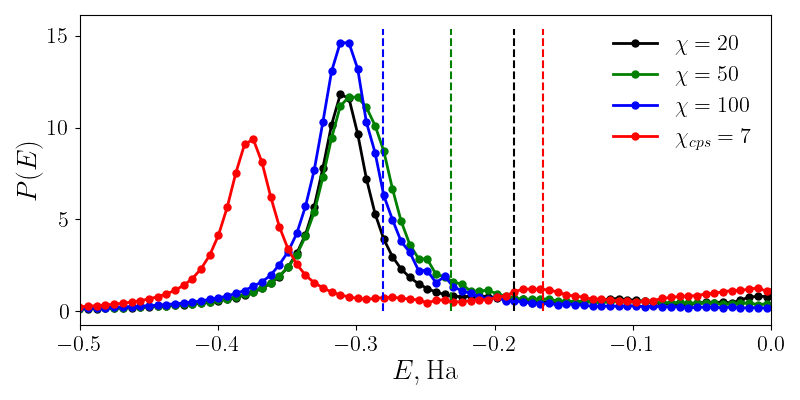}
		\caption{Energy distributions for MPS states at different bond dimensions for the Fe$_4$S$_4$ core in the active space. Energy appears to be a misleading guide for state quality.}
		\label{fig:fe4s4-edists}
	\end{figure}

	The main conclusion to be drawn from \cref{fig:fe4s4-edists} is that even though simple product states have low overlap with the low-energy subspace, it is generally possible to prepare a relatively cheap, relatively high-quality initial state (e.g.~the $\chi_{cps} = 7$ state) for this system with only moderate additional effort. While this particular state is generated from an expensive classical calculation, it retains a considerable weight in the low-energy sectors post-compression, further suggesting that an expensive classical calculation followed by compression could be a good method to obtain a \textit{cheap}-to-implement high-quality state. 
	
	These last two example molecules together suggest that preparing an inexpensive, good-quality state is possible for molecules with transition metal centers. At the same time, such systems are known to be challenging cases for classical computational methods \cite{larsson2022chromium, lee2023evaluating}. The combination of these facts implies that these systems are good candidates for Goldilocks systems, and motivates their further study for quantum computing applications.
	This conclusion is uniquely enabled by the energy distribution picture and data: the overlap metric would not give much insight into the relative quality of the states we considered here.
	
	\section{Conclusions}
	We have introduced a complete workflow for preparing initial states for quantum chemistry. Our results target a critical component of quantum algorithms for simulating chemical systems, which is essential to elucidate the potential for quantum advantage. Key technical contributions of this work include a state-of-the-art quantum algorithm for preparing states expressed as sums of Slater determinants, methods to construct approximate energy distributions for assessing state quality, and identification of coarse quantum phase estimation (QPE) as a leading technique for refining initial states and addressing the leakage problem. 
	All calculations were carried out with the publicly available software library Overlapper \cite{overlapper}. 
	We demonstrate the applicability and usefulness of our initial state preparation procedure with several numerical experiments on challenging molecules.
	
	Our work indicates that it is worthwhile to employ advanced techniques for state preparation beyond simplistic approximations like the Hartree-Fock state. Quantum energy estimation algorithms such as QPE already incur a considerable cost, leaving a large budget available for spending computational resources to prepare better initial states. This budget should be utilized since improved initial states lead to higher probabilities of observing low-energy estimates, resulting in fewer repetitions of the energy estimation algorithm and an overall reduced cost. Our optimized technique for implementing sums of Slater determinants was designed precisely to enable the use of sophisticated approximate ground states such as those obtained from semistochastic heat-bath configuration interaction (SHCI) and the density matrix renormalization group (DMRG) methods, which we identify as leading strategies for initial state preparation.
	
	The energy distribution approach that we propose suggests a rethinking of initial state preparation for quantum chemistry. It provides a computationally-tractable method for assessing and comparing the quality of initial states in a reference-free way; this is out of reach when computing overlaps with the true ground state, which is typically unknown. Energy distributions also help to shape our understanding of the prospects for quantum advantage: since the goal of QPE is to improve the energy estimates associated to the initial state, we can use approximate energy distributions to reason about the extent to which this is possible. We employ this perspective to propose the concept of Goldilocks systems: molecules where the quality of the initial state is neither too high nor too low. This means that two conditions are met: (i) the difference between the best classical estimate and the true ground-state energy is large enough to leave room for improvements, and (ii) the quality of the initial state is sufficiently high to support a considerable probability of observing such improvements.
	
	Numerical experiments support these findings. We observe that it is possible to use energy distributions to infer quality of different initial states, for example in cases where the expectation values of the energy are very similar. This is evidence that energy can be a problematic proxy for state quality. Our studies also suggest that molecules with transition metals in non-equilibrium geometries are potentially Goldilocks systems, and therefore a quantum advantage in ground-state energy estimation could be possible.
	
	Future work may focus on further optimizing quantum algorithms for implementing classical wavefunctions, and more generally, on further improving the proposed workflow. Of particular interest are quantum algorithms for refining initial states obtained from classical methods, which have not received much direct attention. It is possible that better methods than coarse QPE, equipped also with performance guarantees, could be discovered. Another direction that can be pursued is to extend our methodology to periodic systems. This is needed for simulating materials, which have many industrial use cases. Finally, it is important to understand how to prepare initial states in circumstances where quantum hardware places restrictions in terms of available qubits and circuit depth, in preparation for the emergence of early fault-tolerant quantum computers.

	\begin{acknowledgments}
		We thank Huanchen Zhai, Yu Tong and Soran Jahangiri for fruitful discussions. We gratefully acknowledge the support and computational resources provided by the BC DRI Group through the Cedar supercomputing cluster and the Digital Research Alliance of Canada. We are also grateful for the use of computational resources of the LISA computational cluster at the Stewart Blusson Quantum Matter Institute. S.F. acknowledges support by Mitacs through the Mitacs Accelerate Program. J.F. acknowledges support from ERC AdG NOQIA; Ministro de Ciencia y Innovacion AEI (Plan Nacional FIDEUA PID2019-106901GB-I00, Plan Nacional STAMEENA) and Fundació Cellex. 
		This work was also partly supported by the Basic Science Research Program through the National Research Foundation of Korea (NRF), funded by the Ministry of Education, Science and Technology (NRF-2021M3H3A1038085, NRF-2022M3H3A106307411, NRF-2023M3K5A1094805, and NRF-2023M3K5A109481311) and Institute for Information \& communications Technology Promotion (IITP) grant funded by the Korea government(MSIP) (No. 2019-0-00003, Research and Development of Core technologies for Programming, Running, Implementing and Validating of Fault-Tolerant Quantum Computing System). J.H. acknowledges Xanadu for hosting his sabbatical year visit.
	\end{acknowledgments}
	
	\appendix
	
	\section{Conversion to sum of Slater formats for all wavefunction methods}
	\label{app:convert-to-sos}
	Converting all wavefunction-based methods explored in this paper to a sum of Slater format requires a number of specialized steps particular to each method.
	
	The CISD wavefunction already comes in the sum of Slater format, so no conversion is required.
	
	The CCSD ansatz is more challenging to convert to the unified sum of Slaters format due to the fact that in principle excitations to all orders are being generated. However, since these decay quickly, in practice going up to second or fourth order in excitations already captures most of the CCSD wavefunction. These can be obtained by Taylor expanding the exponential to the appropriate order and collecting like terms for excitations: in this way, coupled cluster amplitudes combine in various ways to become CI coefficients. 
	
	CASCI wavefunctions merely need to be padded to the full space with the occupied orbitals, which makes the conversion of these wavefunctions to the sum of Slaters format almost immediate. The same applies to MRPT wavefunctions.
	
	Being one of our standard formats, the MPS does not require form conversion. However, for the purpose of comparison with the other methods, it is also possible to start from an MPS and compute the equivalent Slater determinant representation of the wavefunction up to a specified tolerance -- a process called reconstruction. A deterministic approach to this involves partial re-summation of the matrix products: the details can be found in Ref. \cite{lee2021externally}. On top of that, to switch to chemist convention of keeping all spin up operators on the left, we evaluate the required parity conversion factor for each determinant.
	
	The SHCI method naturally returns the wavefunction as a sum of Slaters, so little conversion is required beyond post-processing the results of the particular package we are employing.
	
	\section{SOS $\leftrightarrow$ MPS transformation}
	\label{app:sos-to-mps-and-back}
	
	In this appendix, we discuss how the two standardized formats, i.e.~SOS and MPS are transformed to each other.

	\medskip 
	
	\noindent{\bf MPS to SOS:}\\ 
	The goal is to calculate the largest coefficients $c(n_1,\ldots,n_N) = \sum_{\alpha_1 \ldots \alpha_{N-1}} A^{n_1}_{1; \ \alpha_1} \ldots A^{n_N}_{N; \ \alpha_{N-1}}$ in a SOS expansion. Based on Appendix A of Ref.~\cite{lee2021externally}, we start from a left canonical form and set a threshhold for keeping terms in the SOS. Partial coefficients such as $c^{(p)}_{\alpha_p}(n_1,\ldots,n_p)=\sum_{\alpha_1 \ldots \alpha_{p-1}} A^{n_1}_{1; \ \alpha_1} \ldots A^{n_p}_{p; \ \alpha_{p-1}\alpha_p}$ are formed and whenever a norm of the partial coefficient $\sum_{\alpha_p} |c^{(p)}_{\alpha_p}(n_1,\ldots,n_p)|^2$ goes below a threshhold, all Slater determinants with the prefix $(n_1,\ldots,n_p)$, i.e. of the form $\ket{n_1,\ldots,n_{a_p},\ldots}$, are dropped from the SOS.
	This way owing to the left canonical form of the MPS, it is ensured that all the terms with coefficients above the threshhold are recovered in the SOS.
	
	\medskip 
	
	\noindent{\bf SOS to MPS:}\\
	For this task, we start with a bond dimension 1 MPS that corresponds to the largest coeeficient Slater determinant in the SOS (could be the Hartree Fock state or not); we make an auxiliary copy of it also. Using MPOs consisting of a number of $c^\dagger$ and $c$ operators, the auxiliary bond dimension 1 MPS is transformed to the Slater determinant with the second largest coefficient. The new auxiliary MPS is added to the main MPS and the procedure goes on until all coefficients are added. Note that the auxiliary MPS remains bond dimension 1, while the bond dimension of the main MPS grows, one can compress the main MPS as more and more terms are added to it.

	\section{Proof of Lemma \ref{lem:unique_signature}}\label{appsec:proof_of_compression_lemma}
	We prove \cref{lem:unique_signature} by induction. To avoid cluttering, we shall replace $\tilde{\nu}_i$ with $\nu_i$. From here onwards, the vectors $\nu_i$ have length $r$. We recall the statement and notations of the lemma: In \cref{eq:U_V_B}, one needs to prove the existence of $2\log D-1$ vectors $u_j$ of length $r$, forming a matrix called $U$, that helps to distinguish the $D$ distinct vectors $\nu_i$ of length $r$ by mapping them to vectors $b_i$ of length $2\log D-1$. Notice $U$ is supposed to be found offline on the classical computer.
	
	We can interpret $U$ as a linear map acting on each $\nu_i$. We need to find a $U : \mbbF_2^r \to \mbbF_2^{2\log D-1}$ such that $U(\nu_i) \neq U(\nu_j) \leftrightarrow \nu_i - \nu_j \not \in \ker U , \forall i \neq j$ . We will efficiently construct such a linear map with the additional property that $\nu_i \not \in \ker U, \forall i$ unless $\nu_i = \bm{0}$; this additional property will help us in proving the next inductive step from the induction hypothesis. In summary, $U$ has to satisfy the following properties:
	\begin{align}\label{eq:properties_ker_U}
		\nu_i - \nu_j \not \in \ker U , \forall i \neq j, \ \ \nu_i \not \in \ker U, \forall i, \text{ unless } \nu_i = \bm{0}.
	\end{align}
	
	\begin{proof}[Proof of \cref{lem:unique_signature}]
		Since $ \log D-1 < \log(D) \implies \dim \mbbF_2^{\log D-1} = 2^{\log D-1}<D$, there are at least $\log D$ many linearly independent vectors among the $\nu_i$'s. Therefore $r \ge \log D$. First assume $\log D \le r \le 2\log D-1$. Without loss of generality, assume $\nu_1, \ldots , \nu_r$ are linearly independent and generate all the $\nu_i$'s. Note that finding these linearly independent generators is an efficient classical algorithm in linear algebra. In this case, we can distinguish the $\nu_i$'s using $r\le 2\log D-1$ many $u_j$'s; we simply choose $U$ to be the $r\times r$ identity matrix. This choice yields $b_i = \nu_i$, with length $r\le 2\log D-1$, fulfilling the same purpose.

		When $r>2\log D-1$, we perform induction on $t \in \mbbN$ where $r=2\log D-1+t$. According to the rank theorem, $\dim \text{Im }U+ \dim \ker U = \dim \mbbF_2^r = r = 2\log D-1+t$. We find $U$ by first constructing a subspace $\mcW$ of dimension $t$, satisfying the same properties in \cref{eq:properties_ker_U}, followed by building a $U$ with kernel equal to $\mcW$. More precisely, we will find $t$ many linearly independent vectors $w_1,\ldots,w_t \in \mcV_r :=\text{span}\langle \nu_1,\ldots,\nu_r \rangle$ that would define such a $\mcW$. Then, by basic linear algebra, there is an efficient classical algorithm that finds linearly independent vectors $u_1,\ldots,u_{2\log D-1} \in \mcV_r$ that satisfy $u_j \cdot w_i = 0, \forall i,j$. Because of their linear independence, a matrix $U$ defined by such $u_j$'s would have rank $2\log D-1$, so $\dim \text{Im } U = 2\log D-1 \implies \dim \text{ker }U = t$. Finally, since $\forall i: w_i \in \ker U$ and $\dim \ker U = \dim \mcW$, it follows $\ker U = \mcW$. Therefore $\ker U$ satisfies \cref{eq:properties_ker_U}, as desired.
		
		\textit{Note. }Going forward, as operations are over the field $\mbbF_2$, we may play loose with subtraction and addition, as $\nu_i-\nu_j = \nu_i+\nu_j$. 
		
		Let us start by proving the base of induction $t=1 \implies r = 2\log D$. We need to find a single vector $w_1$ such that $\nu_i \neq w$ and $\nu_i - \nu_j \neq w$. The number of distinct vectors in the set $\{\nu_i, \nu_i-\nu_j\}_{i,j}$ is at most $D+\binom{D}{2}= (D^2+D)/2\le 2^{2\log D-1}+2^{\log D-1}< |\mbbF_2^r| = 2^r = 2^{2\log D}$. Therefore there exists $w \in \mbbF_2^r - \{\nu_i, \nu_i-\nu_j\}_{i,j}$, and this vector can be found after a search over $(D^2+D)/2+1$ vectors picked from $\mbbF_2^r$. Thus, $w_1$ for the base of induction can be found efficiently.
		
		For the induction step, without loss of generality assume $\nu_1, \ldots, \nu_{2\log D-1+t}$ are all linearly independent and generate the rest of the $\nu_i$'s (we note again that finding these generators can be done efficiently). By induction hypothesis $ w_1, \ldots , w_{r-1}\in \mcV_{r-1} = \text{span} \langle \nu_1, \ldots , \nu_{r-1} \rangle$ form a desired subspace $\mcW_{r-1}$ for the previous induction step. Note that clearly $\mcV_{r-1} \subset \mcV_{r} = \text{span} \langle \nu_1, \ldots , \nu_r \rangle$.
		
		We can partition the set of all vectors $\{\nu_i\}_{i=1}^D$ into three subsets: (1)  $\mcM := \{\nu_i \ | \ \nu_i \in \mcV_{r-1}\}$, which elements will be referred to  as $m_i$, (2) the single element subset $\{\nu_r\}$, and (3)  $\mcN := \{\nu_i \ | \ \nu_i \in \mcV_{r} - \mcV_{r-1}\}$. The latter will have vectors that look like $\nu_i = m_i'+\nu_r$ where $0 \neq m_i'\in \mcV_{r-1}$. Because of the partitioning, $|\mcM| + 1 + |\mcN| = D$. Note any future use of $m_i,m_i'$ will refer to a $\nu_i$ inside $\mcM,\mcN$ respectively. We emphasize that $\mcM,\mcN$ are sets and not necessarily a linear subspace.
		
		We would like to invent a new set of $D$ vectors with rank $r-1$, so that we can apply the induction hypothesis. To do so, let us replace $\nu_r$ with some $l \in \mcV_{r-1}$, and similarly substitute every $\nu_r$ in the linear expansion of any $\nu_i = m_i' +\nu_r \in \mcN$, meaning $\nu_i$ becomes $l+m_i'$. This vector $l$ needs to satisfy some properties: 
		\begin{align}\label{eq:l_properties}
			&l \neq \bm{0} , m_i'+l \neq \bm{0}, \\ \label{eq:l_properties2}
			&l \neq m_j,  m_i'+l \neq m_j
		\end{align}
		The first two conditions ensure that after replacement, we do not obtain any zero vector. The second line ensures that we do not obtain any repeated vector. All these conditions amount to $l \not \in  \{\bm{0}, m_i', m_j, m_j+m_i' \}$, the size of this set being (at most) $1 + |\mcN| + |\mcM| + |\mcM|\cdot |\mcN|$. We recall $1 + |\mcN| + |\mcM| = D$, so the size is $\le D + (D-1-|\mcN|)|\mcN| \le D + D^2/4 \le 2^{\log D} + 2^{2\log D-2} < 2^{r-1}=2^{2\log D-t-2}$ as $t>1$. Hence, there exists $l \in \mcV_{r-1}$ that satisfies \cref{eq:l_properties,eq:l_properties2}.

		Now the induction hypothesis for the new set of $D$ vectors apply, since the rank has clearly been decreased by one to $r-1 = 2\log D-1 +t-1$. Therefore, there exists a subspace $\mcW = \text{span}\langle w_1,\ldots,w_{t-1}\rangle \subset \mcV_{r-1}$ satisfying \cref{eq:properties_ker_U} for this new set of vectors. Now let us bring back $\nu_r$ by undoing the replacement by $l$. After this change, we need to verify that $\mcW$ still satisfies \cref{eq:properties_ker_U}, and then, in order to finish the proof, extend $\mcW$ by a vector $w_t$ while satisfying said properties. We verify the properties as follows:
		\begin{itemize}
			\item We first check that $\nu_i, \nu_i+\nu_j \not \in \mcW$. Note that $\nu_i \not \in \mcW$ needs to be checked only for $\nu_r$ and $\nu_i \in \mcN$ (as they are the only ones impacted by bringing back $\nu_r$). For both, this is in fact obvious as $(\{\nu_r\} \cup \mcN) \subset \mcV_r -  \mcV_{r-1} \implies (\{\nu_r\} \cup \mcN) \cap \mcV_{r-1} = \emptyset$ while $\mcW \subset \mcV_{r-1}$.
			\item For the property $\nu_i+\nu_j \not \in \mcW$, this needs to be checked only when at least one of $\nu_i,\nu_j$ is inside $(\{\nu_r\} \cup \mcN)$. 
			\begin{itemize}
				\item Assume that $\nu_i \in (\{\nu_r\} \cup \mcN)$ and $\nu_j = m_j \in \mcM$. If $\nu_i=\nu_r$, we need to show $\nu_r+m_j \not \in \mcW$ and for $\nu_i \in \mcN$, we need to prove $m_i'+\nu_r+m_j\not \in \mcW $. Both these cases, due to $\nu_r$, are outside of $\mcV_{r-1}$ and $\mcW \subset \mcV_{r-1}$.
				\item Assume that $\nu_i,\nu_j  \in (\{\nu_r\} \cup \mcN)$. Then we need to check $\nu_r + m_j' + \nu_r = m_j'\not \in \mcW$, and also $m_i' + \nu_r + m_j' + \nu_r = m_i'+m_j'\not \in \mcW$. However by induction hypothesis, we already know that $l + m_j' + l = m_j'\not \in \mcW$  and  $l+m_i'+l+m_j' = m_i'+m_j' \not\in \mcW$, so this is also guaranteed.
			\end{itemize} 
		\end{itemize}
		
		Finally, we need to find a new $w_t$ to add to $\mcW$ while preserving its properties. Let us define $w_t = \nu_r +l$ and let $\mcW' = \mcW \oplus w_t \subset \mcV_r = \mcV_{r-1} \oplus \nu_r$. To prove that $\mcW'$ satisfies \cref{eq:properties_ker_U}, one needs to verify that $\nu_i \not \in \mcW' - \mcW$ or $\nu_i-\nu_j \not \in \mcW' - \mcW$, as $\mcW$ has already been shown to satisfy said properties. If $\nu_i \in \mcW' - \mcW$ or $\nu_i-\nu_j  \in \mcW' - \mcW$ then $w_t$ must be `involved': 
		\begin{itemize}
			\item For $\nu_i \in \mcW' - \mcW$, we must have $\nu_i = w_t+w = \nu_r+l + w $ for some $w \in \mcW$, in which case $\nu_i-l = \nu_r + w$. However, the latter is inside $\mcV_r-\mcV_{r-1}$. Therefore, since $l \in \mcV_{r-1}$, we have $\nu_i \in \mcV_r- \mcV_{r-1}$. So $\nu_i \in (\{\nu_r\} \cup N)$. If $\nu_i = \nu_r$ then $l = w \in \mcW$, which violates our induction hypothesis. If $\nu_i= m_i' + \nu_r \in N$  then $m_i'+l = w \in \mcW$, which again violates the construction of $\mcW$.
			\item For $\nu_i-\nu_j =  \nu_r+l + w$ for some $w \in \mcW$,  exactly one of $\nu_i$ or $\nu_j$ must be inside $(\{\nu_r\} \cup N)$. Without loss of generality assume $\nu_i = \nu_r$ or $\nu_i= m_i'+\nu_r$. Then this simplifies to $l-\nu_j = w \in \mcW$ or $m_i'+l-\nu_j = w \in \mcW$, both violating the induction hypothesis for $\mcW$.
		\end{itemize}
		This shows $\mcW'$ satisfies \cref{eq:properties_ker_U} and finishes the induction. The significant cost in each inductive step is the search to find $l$, taking $O(D^2/2+D)$ steps. 
		
		\textbf{Resource estimation. }The total complexity is found by applying this for each induction step, thus $O(tD^2)$. Note that $t \le \min(2N,D)-2\log D+1$, with equality when all the vectors $\nu_i$ are linearly independent; so the total cost of the classical algorithm used to find $U$ is at most $O(D^2(\min(2N,D)-2\log D+1))$. It should be noted that the search process can be fully parallelized, using all cores on an available machine, and given the nature of this search, the expected runtime could be much less.
	\end{proof}
	
	\section{Trading off Toffolis with qubits in the SOS algorithm}\label{appsec:trading_off_tof_w_qubits}
	As explained in the main text, trading off Toffolis with qubits can be done by using an alternative version of QROM. This variant, which we shall call \ssdirty \cite[Fig. 1d]{low2018trading}, has a parameter $\lambda$ that allows for trading off qubits with Toffolis. For a QROM loading $L$ many data points $\ket{x_i}$,  indexed by $i = 1,\ldots , L$ and of size $c$, the trade-off $\lambda \in [1,L]$ can be applied to change the Toffoli complexity from $O(L)$ to $O(L/\lambda +\lambda c)$ while increasing the uninitialized (so-called dirty) qubit cost to $O(\lambda c)$. Notice that the volume cost stays as $O(Lc)$, although technically, as we traded gates with dirty qubits, this volume is not a clean volume, so it is an overall improvement. To keep our discussion focused on the novelties and following the convention in previous resource estimations \cite{berry2019qubitization}, we will select $\lambda = \sqrt{L/c}$ in our applications. This strikes a balance in the trade off, using `equally' many ($O(\sqrt{Lc})$) Toffolis as dirty qubits.
	
	The first expensive QROM is employed when outputting the system register in \cref{eq:first_qrom_sys_register}. Using the \ssdirty variant with Toffoli cost $2D/\lambda + 8 (2N)\lambda$ with $\lambda = \sqrt{2D/16N}$, this QROM Toffoli cost can be lowered to $2\sqrt{32ND}$ while also using $\sqrt{32ND}$ dirty qubits.
	
	We also could have chosen to use \ssdirty QROM to flip the register $\ket{i}$ using $\ket{b_i}$ in
	\begin{align}\label{eq:i_v_b_superposition}
		\sum_{i=1}^D \alpha_i\ket{i}\ket{\nu_i}_v\ket{ b_i}_b,
	\end{align}
	where we have denoted the registers by subscripts. We need to employ the inverse $O^\dagger$ of the \ssdirty operator $O \ket{b_i}_b\ket{0}^{\otimes \log D} = \ket{b_i}_b\ket{i}$. The naive implementation of $O$ would read $2\log D-1$ qubits of $b_i$, therefore its optimized Toffoli cost would scale as $O(\sqrt{D^2 \cdot \log D}) = O(D \sqrt{\log D})$. However, know that we only have $D \sim 2^{\log D}$ many $b_i$'s and we would like to exploit this fact, achieving a Toffoli cost that is sublinear in $D$. 
	
	First we note that we have knowledge of the value of $b_i$ as we computed them classically. Let $\lambda= \lceil \sqrt{D}\rceil$. Let us now order $b_{l_1} < \cdots < b_{l_{D}}$. We compute:
	\begin{align}
		\ket{b_j}\ket{0}^{\otimes \lceil \log D/2\rceil} \to \ket{b_j}\ket{f(j)}
	\end{align}
	where $f(j)$ is the unique index such that $b_{l_{\lambda (f(j)-1)}} < b_j < b_{l_{\lambda f(j)}}$, where for $\lambda f(j)>D$, we set $i_{\lambda f(j)} = D$. This computation requires comparing $b_j$ to  $\lceil D/\lambda \rceil$ many other $b_i$'s. This means a Toffoli cost of $(2\log D-1+2\log D-1)$ (the comparator cost of two $2\log D-1$ bits integers) multiplied by the $\lceil \frac{D}{\lambda}\rceil $ comparisons that we have to make. More precisely, starting from $b_{q\lambda}$ for $q=1$, we compare $b_j$ with $b_{q\lambda}$ stored in an auxiliary register, and if $b_j > b_{q\lambda}$, we store $\ket{1}$ in another auxiliary register and otherwise $\ket{0}$. Then we add the value of that register to another register holding $\ket{q}$ which will eventually become $\ket{f(j)}$. 
	This summation itself costs $\lceil \log D/2\rceil+1$ Toffolis as the size of the register holding $q$ is $\lceil \log D/2\rceil$ (the qubit cost is effectively zero, as any ancilla used is immediately liberated without Toffolis). As this summation must be performed for all $q \in \{1, \ldots,\lceil \frac{D}{\lambda}\rceil\}$, we have an additional $\lceil \log D/2\rceil \cdot \lceil \frac{D}{\lambda} \rceil $ to account for. In total, the cost for this computation is $(4\log D-2+\lceil \log D/2\rceil)\lceil \frac{D}{\lambda} \rceil \sim \frac{9\log D\sqrt{D}}{2}$, and we note that its (later) uncomputation can be done with Clifford gates. After this, our state becomes 
	\begin{align}
		\sum_{i=1}^D \alpha_i\ket{i}\ket{\nu_i}_v\ket{b_i}_b\ket{f(i)}
	\end{align}
	Next, we read the register $f(i)$ using a \textsc{Select} QROM with cost $2^{\log D/2}$ and output the $\lambda$-block of $b_j$'s in which we know $b_i$ lives in, i.e. : $\ket{b_{l_{\lambda (f(j)-1)}+1}, \ldots, b_{l_{\lambda f(j)}}}$.
	\begin{align}
		\sum_{i=1}^D \alpha_i\ket{i}\ket{\nu_i}_v\ket{b_i}_b\ket{f(i)}\ket{b_{l_{\lambda (f(i)-1)}+1}, \ldots, b_{l_{\lambda f(i)}}}
	\end{align}
	Notice the significant clean qubit cost $(2\log D-1)\lambda \sim (2\log D-1)\sqrt{D}$. The uncomputation of this step is later done with Clifford gates. CNOT-ing the register $b_i$ into the block yields:
	\begin{align}
		\sum_{i=1}^D &\alpha_i\ket{i}\ket{\nu_i}_v\ket{b_i}_b\ket{f(i)}\otimes \nonumber \\
		&\ket{b_{l_{\lambda (f(i)-1)}+1}\oplus b_i, \ldots, b_{l_{\lambda f(i)}}\oplus b_i}
	\end{align}
	By using $\lambda$ many $(2\log D-1)$-MCCNOTs, with total Toffoli cost $(2\log D-1)\lambda$, we compute the index $1\le g(i) \le \lambda$ for which $b_{l_{\lambda(f(i)-1)+g(i)}} = b_i$. More precisely, we hold in an auxiliary register the index $\ket{k}$ at which we are implementing the MCCNOT, and using Toffolis, bit by bit for every $\lceil\frac{\log D}{2}\rceil$ bits, add that to the register designed for holding $g(i)$ controlled on the result of the $(2\log D-1)$-MCCNOT. This requires an additional $\lceil \log D/2 \rceil$ Toffolis and ensures that only the true $k=g(i)$ is added to the register designed for holding $g(i)$. Therefore, the total Toffoli cost of this step is $(2\log D+\lceil \log D/2 \rceil -1)\lambda \sim \frac{5\log D\sqrt{D}}{2}$. We now have the following state
	\begin{align}
		\sum_{i=1}^D &\alpha_i\ket{i}\ket{\nu_i}_v\ket{b_i}_b\ket{f(i)}\otimes \nonumber \\
		&\ket{b_{l_{\lambda (f(i)-1)}+1}\oplus b_i, \ldots, b_{l_{\lambda f(i)}}\oplus b_i} \ket{g(i)}
	\end{align}
	Notice the register $\ket{f(i),g(i)}$ has $2\lceil \frac{\log D}{2}\rceil \sim \log D$ qubits, and determines $i$ uniquely. Tracing this back to our original goal of distinguishing the $D$ many $\nu_i$'s, this is the ideal we can hope for, as we have computed $D$ many $\log(D)$-bits integers $\ket{f(i),g(i)}$ that are distinct. The rest can be done using the inverse $O^\dagger$ of a \ssdirty QROM which computes 
	\begin{align}
		O\ket{f(i),g(i)} \ket{0}^{\otimes \log D}    = \ket{f(i),g(i)} \ket{i}
	\end{align}
	Notice that we have access to where $b_i$ is in the list $b_{l_1}<\ldots <b_{l_D}$, and therefore can classically compute the values $f(i),g(i)$. Hence the QROM has the required classical lookup data. This \ssdirty QROM has Toffoli  cost $2\sqrt{16\log(D)2^{\log D}} < 2\sqrt{32\log(D)D}$ with dirty qubit cost $<\sqrt{32\log(D)D}$. Note that these dirty qubits are already available using the $\lambda-$block which has $(2\log D-1)\sqrt{D}$ qubits.
	
	\textbf{Resource estimation. }Overall, the clean qubit cost increases to $(2\log D-1)\sqrt{D}$, and an additional dirty qubit cost $\sqrt{32ND}$. The clean qubit cost may be improved if one can find a way to compute $g(i)$ without outputting the entire $\lambda-$block. While the dirty qubit cost can be avoided if we do not apply \ssdirty QROM in \cref{eq:first_qrom_sys_register}. In practice, the dirty qubit cost, though significantly larger than the clean ($5\log D$) qubit cost in our previous method, may be actually available if a future computation (such as qubitization) to simulate the evolution of the Hamiltonian requires that much qubit. The Toffoli cost changes to $\min(2\sqrt{32ND},D) + \frac{9\log D\sqrt{D}}{2}+\frac{5\log D\sqrt{D}}{2}+2^{\log D/2}+2\sqrt{32log(D)D} \sim \min(2\sqrt{32ND},D) + (7\log D+2\sqrt{32\log D})\sqrt{D}$. In case $D$ is chosen from the minimum, the dirty qubit cost $\sqrt{32ND}$ is lifted.
	
	\textbf{Is the algorithm optimal? }Our algorithm is general in the sense that it simply assumes a given set of amplitudes $\{\alpha_i\}$ and bitstrings $\{\nu_i\}$, with no assumption on the nature of either of the two sets. Furthermore, a dimensionality argument can show that the compression in \cref{lem:unique_signature} of $\nu_i$ to $2\log D -1$ bits is very likely to be tight, and we conjecture that the volume cost can not be asymptotically lower than $\tilde{O}(D)$, and more strongly, $O((\log D)^2D)$. We remark that this conjecture is for creating the superposition $\sum_{i=1}^D \alpha_i \ket{\nu_i}$, without any junk register, i.e. $\sum_{i=1}^D \alpha_i \ket{\nu_i}\ket{\text{junk}_i}$ is not acceptable. The only `approximation' allowed is in the amplitudes, and the distance to the state should be on the order of chemical accuracy. Both These restrictions are necessary as this preparation problem concerns the system register, and not, say, a PREP state in a qubitization protocol derived from an LCU, which \textit{can} include a junk register, and hence why preparation methods such as coherent alias sampling \cite{berry2019qubitization} can be employed in that instance.
	
	One wonders if the approach in \cite{tubman2018postponing} in iteratively generating the superposition can be combined with ours. This would involve ordering the states $\nu_i$ and compressing $\nu_1, \ldots, \nu_l$ for each $1\le l \le D$. Assuming that for each $l$, a compression of the bitstrings $\nu_1, \ldots, \nu_l$ to a length $k_l$ is possible (and we know $k_l \le 2\lceil \log l\rceil -1$), then the Toffoli cost of generating the superposition is $\sum_{l=1}^D (k_l-1)$. This is upper bounded by $\sum_{i=1}^{\lceil \log D\rceil -2} (2i-1)(2^i-1) +  (2\lceil \log D\rceil -1)(D-2^{\lceil \log D \rceil -1})$ and can be seen to be smaller than our own cost $(2\lceil \log D \rceil+3)D$. 
	However, this method also involves a rotation \textit{at each step} for $1\le l\le D$, and even assuming access to a gradient state, the overall cost of these rotations is $(c-3)D$ Toffolis, where $c$ is the required bit precision. Note that we also have to take into account the accuracy $c$ for our rotations in generating the superposition in \cref{eq:superposition_enumerator}, however the associated cost is simply $(c-3)\log D$, and according to Lemma E.1 in \cite{zini2023quantum}, $c$ must satisfy $c > \log ( \log(D) \pi / \epsilon)$, where $\epsilon$ is chemical accuracy. Crucially, $c$ is double logarithmic in terms of $D$, because there are only $\log D$ rotations performed in generating \cref{eq:superposition_enumerator}. 
	This is not the case when using the approach in \cite{tubman2018postponing}, and for the overall error to be under chemical accuracy, we need $c > \log(D\pi/\epsilon)$, therefore making $c$ logarithmic in $D$. Still, one could see the total cost $\sum_{i=1}^{\lceil \log D\rceil -2} (2i-1)(2^i-1) +  (2\lceil \log D\rceil -1)(D-2^{\lceil \log D \rceil -1}) + (\log(D\pi/\epsilon)-3)D$ to be competitive within a constant factor of two with $(2\lceil \log D \rceil+3)D$, but clearly, it does not lead to a significant constant factor cost reduction.
	Nevertheless, this is a worst case complexity analysis, and if $k_l$'s are significantly smaller, one might see benefits of this combination of the two methods. This could potentially be the case for a particular range of $D$ and bitstrings $\nu_i$'s which have some common structure, such as being an excitation or two away from a reference state, and we leave that for future works.
	
	\section{Details of resource estimation for MPS implementation}\label{app:mps_implementation_details}

	In this appendix we discuss the details of the MPS cost estimation presented in \cref{sec:mps_implementation_main}. In particular we focus on the ancilla qubit and Toffoli costs of implementing $G[j]$ defined in \cref{eq:G_def_tensors_grpah}.
	
	We represent the $G$ operations as:
	\begin{equation}\label{eq:G_dirac_form}
		G[j] = \sum_{\alpha_{j-1}} \left( \ket{u_{\alpha_{j-1}}} \bra{\alpha_{j-1},0} \right) + \ldots,
	\end{equation}
	with
	\begin{equation}
		\ket{u_{\alpha_{j-1}}} = \sum_{\alpha_j,n_j} A^{n_j}_{j; \ \alpha_{j-1} \alpha_j} \ket{\alpha_j, n_j}.
	\end{equation}
	In the states $\ket{\alpha_j, n_j}$ the first and second arguments show the ancillae and system qudit indices respectively. 
	Note that the form of \cref{eq:G_dirac_form} follows from the defintion of $G$ in \cref{eq:G_def_tensors_grpah}.
	Each unitary is synthesized using a series of Householder reflections with the addition of a single ancilla qubit as follows:
	\begin{equation}\label{eq:G_as_householder_reflections}
		\ket{0}\bra{1} \otimes G[j] + \ket{1}\bra{0} \otimes G[j]^\dagger = \prod_{\alpha_{j-1}=1}^{\chi_{j-1}} (1-2 P_{\alpha_{j-1}})
	\end{equation}
	where a total of $\chi_{j-1}$ reflections are used, whose projectors are defined through an auxiliary state $\ket{w_{\alpha_{j-1}}}$:
	\begin{equation}
		P_{\alpha_{j-1}} = \ket{w_{\alpha_{j-1}}} \bra{w_{\alpha_{j-1}}}, 
	\end{equation}
	with
	\begin{equation}\label{eq:w_ket_W_op}
		\begin{aligned}
			\ket{w_{\alpha_{j-1}}} &= \ket{1}\otimes \ket{\alpha_{j-1},0} - \ket{0} \otimes \ket{u_{\alpha_{j-1}}} \\
			&= W_{\alpha_{j-1}} \ket{0} \otimes \ket{0,0}.
		\end{aligned}
	\end{equation}
	On the second row, we have defined the operator $W_{\alpha_{j-1}}$ that prepares $\ket{w_{\alpha_{j-1}}}$, also in the state $\ket{0} \otimes \ket{0,0}$ the 0's correspond to the reflection ancilla, MPS ancillae and the system qudit respectively.
	
	With the operator $W_{\alpha_{j-1}}$, each of the reflections in \cref{eq:G_as_householder_reflections} can be written as:
	\begin{equation}
		1-2P_{\alpha_{j-1}} = W_{\alpha_{j-1}} \, \left[ 1-2 \ket{0,0,0}\bra{0,0,0}\right] \,  W_{\alpha_{j-1}}^\dagger,
	\end{equation}
	and thus the cost of implementing each reflection is twice the cost of $W_{\alpha_{j-1}}$ plus the cost of the simple reflection $\left[ 1-2 \ket{0,0,0}\bra{0,0,0}\right]$. As a result, we need to also evaluate the cost of $W_{\alpha_{j-1}}$, which we synthesize as follows: we first define an operator that prepares the state $\ket{u_{\alpha_{j-1}} }$:
	\begin{equation}
		\ket{u_{\alpha_{j-1}}} = V_{\alpha_{j-1}}\ket{0,0}.
	\end{equation}
	It can be seen that $W_{\alpha_{j-1}}$ as below can serve to satisfy the definition of $\ket{w_{\alpha_{j-1}}}$ in \cref{eq:w_ket_W_op}:
	\begin{equation}\label{eq:W_clifford_nonclifford}
		W_{\alpha_{j-1}} = \left(\bar{C}V_{\alpha_{j-1}} \right) \left( C_{\alpha_{j-1}} \right) \left( (ZH) \otimes \mathbb{I} \right),
	\end{equation}
	where $ZH$ (a Pauli $Z$ and a Hadamard) acts on the reflection ancilla, $C_{\alpha_{j-1}} $ is a product of CNOTs controlled on the reflection ancilla to prepare the state $\frac{1}{\sqrt{2}}\left[\ket{0,0,0}- \ket{1,\alpha_j,0} \right]$. With $\bar{C}V_{\alpha_{j-1}}$ controlled negatively on the reflection ancilla, one can check that $\ket{w_{\alpha_{j-1}}}$ is prepared up to a phase.
	
	We are interested in the Toffoli cost of implementation and list all sources of Toffoli cost below:
	\begin{itemize}
		\item Simple reflections $\left[ 1-2 \ket{0,0,0}\bra{0,0,0}\right]$: a number of $\chi_{j-1}$ of them is required for $G[j]$.
		\item Operations $\bar{C}V_{\alpha_{j-1}}$: these are required for creating $W_{\alpha_{j-1}}$ and are the only source of non-Clifford gates in their synthesis. A total of $2 \chi_{j-1}$ of such operators is required for $G[j]$.
	\end{itemize}

	First we note that the reflection $\left[ 1-2 \ket{0}^{\otimes \nu}\bra{0}^{\otimes \nu}\right]$ can essentially be thought of as a multi-controlled Z operation, and thus can be implemented using a circuit such as the one shown in figure 4.10 of \cite{nielsen2010quantum}. However, the second half of the circuit, i.e.~the uncomputing part can be done without Toffolis and in fact using measurements and Clifford gates as shown in figure 3 of Ref.~\cite{gidney2018halving}. This makes the total number of Toffoli gates and ancillae equal to $\nu-1$.

	Next, we discuss how $\bar{C}V_{\alpha_{j-1}}$ can be implemented and estimate the required resources. For this we first discuss the implementation of $V_{\alpha_{j-1}}$.
	$V_{\alpha_{j-1}}$ prepares the state:
	\begin{equation}
		\ket{u_{\alpha_{j-1}}} = V_{\alpha_{j-1}}\ket{0,0},
	\end{equation}
	which is defined as:
	\begin{equation}
		\ket{u_{\alpha_{j-1}}} = \sum_{\alpha_j,n_j} A^{n_j}_{j; \ \alpha_{j-1} \alpha_j} \ket{\alpha_j, n_j}.
	\end{equation}
	First, we take the above subspace of interest to consist of $\nu$ qubits.
	The preparation of a generic state as $\ket{u_{\alpha_{j-1}}}$ can be done using the methods discussed in Ref.~\cite{low2018trading}; the state is carved qubit by qubit in $\nu$ steps; in each step a single qubit rotations on one qubit being controlled on all the previous entries is performed and with consecutive application of this procedure all the correct probabilities for bitstrings are reproduced at the end; one last multicontrolled single qubit rotation is required to recover the complex phases corresponding to the components of the state in question (see page 3 of \cite{low2018trading} for details). Thus a total of $\nu+1$ of such single qubit rotations are required for reproducing the state. 
	
	The rotation will be performed with the method given in \cite{gidney2018halving}, where access to a phase gradient state $2^{-b/2}\sum_{k=0}^{2^{b}-1} e^{-2\pi i k / 2^b}\ket{k} $ is assumed. $b=\log(1/\delta_r)$ is the number of digits in the binary representation of the rotation angle and thus $\delta_r$ is the error in rotation.
	The Toffoli cost of each single qubit rotation is given by $b+O(1)$ \cite{gidney2018halving}; note that this also means we need an additional $\log(1/\delta_r)$ additional qubits to store the phase of each single qubit rotation.

	Considering first the \textsc{Select} variant of implementation in \cite{low2018trading}, we now discuss the cost of control operations for the above single qubit rotations. Since multicontrolling over a sequence of $0,1,2,\ldots,\nu$ qubits is required to store the respective rotation angles, we will respectively have a sequence of Toffoli costs of $2^0-1,2^1-1,2^2-1,\ldots,2^\nu-1$ according to \cite{babbush2018encoding} (see e.g.~figure 7 therein). Note that we are interested in implementing $\bar{C}V_{\alpha_{j-1}}$, and \cite{babbush2018encoding} also considers a controlled operation for the above cost.
	After each single qubit rotation, the qubits storing the rotation angle should be uncomputed and this adds a multiplicative factor of 2.
	As a result, for the \textsc{Select} variant we have a total Toffoli cost equal to
	\begin{equation}
		2^{\nu+2} +  \nu b.
	\end{equation}
	Where we have dropped an additive $-\nu$ term in the sum as it is subdominant.
	
	In our particular case of interest, i.e.~synthesis of $G[j]$, we have a number of reflections shown in \cref{eq:G_as_householder_reflections}, equal to $\chi_{j-1}$. On the other hand the Hilbert space over which each of the relections acts is $\chi_j d$ and as a result, $N$ in the above trwatments should be $2^\nu = \chi_j d$. This means that in the \textsc{Select} variant, the total Toffoli cost reads:
	\begin{equation}
		\chi_{j-1} \left[ 8\chi_j d + b \log(\chi_j d) + \log(\chi_j d) \right],
	\end{equation}
	
	Next turning to the other variant \textsc{SelSwapDirty} of Ref.~\cite{low2018trading}, which is capable of reducing the Toffoli cost if dirty qubits are available; we saw above that a number $b$ qubits is required for storing the rotation angles, however with the addition of a number $\lambda b$ dirty qubits, we can use this variant of the algorithm. A total of $\nu+b$ extra clean qubits are also required (excluding the ancillae required for performing single qubit rotations like the phase gradient state); note that this is the same number as the \textsc{Select} variant.
	Moreover, for \textsc{SelSwapDirty}, one needs to perform swaps also which will add to the total Toffoli cost.
	It is straightforward to see that the Toffoli gate cost in this case reads:
	\begin{equation}
		2\frac{2^{\nu+2}}{\lambda} + 4 \cdot 2\lambda \nu b + \nu b.
	\end{equation}
	The first term corresponds to multi-qubit controls, the second term swaps and the third term single qubit rotations. The factors 2 and 4 in the first and the second terms appear because \textsc{Select} and \textsc{Swap} need to be done twice and four times in \textsc{SelSwapDirty} (See figure 1d of \cite{low2018trading}). The factor of 2 in the second term comes from uncomputing the rotation angles.
	
	As is discussed in \cite{low2018trading}, it is best for Toffoli gate count to be $\lambda = O(\sqrt{2^\nu})$, but we will keep it unspecified for the rest of the discussion.

	Gathering all the above costs together for synthesizing $G[j]$ in \cref{eq:G_dirac_form}, we see that the Toffoli cost reads using the \textsc{SelSwapDirty} variant:
	\begin{equation}
		\chi_{j-1} \left[ 8\frac{\chi_j d}{\lambda} + 8 \lambda b \log(\chi_j d) + b \log(\chi_j d) + \log(\chi_j d) \right]
	\end{equation}
	
	In total assuming a number $N$ of qudits in the physical system, the total cost will be the sum of the above; asymptotically and with using $\chi$ selctively for all bond dimensions, the dominant Toffoli cost can be written as $O(N \chi^{3/2})$.

	\section{Edgeworth series terms}\label{app:edgeworth series}
	\begin{table*}[!t]
		\centering
		\begin{tabular}{|c|l|}
			\hline
			$s$ & Edgeworth term \\ \hline
			1 & $(\kappa_3/6) \, \He_3(x)$ \\ \hline
			2 & $(\kappa_4/24) \, \He_4(x) + (\kappa_3^2/72)   \,  \He_6(x) $ \\ \hline
			3 & $(\kappa_5/120) \, \He_5(x) + (\kappa_4 \kappa_3/144)   \,  \He_7(x) + (\kappa_3^3/1296) \, \He_9(x) $ \\ \hline
			4 & $(\kappa_6/720) \, \He_6(x) + (\kappa_5 \kappa_3/720 + \kappa_4^2/1152)   \,  \He_8(x) +  (\kappa_4 \kappa_3^2/1728)   \,  \He_{10}(x)  + (\kappa_3^4/31104) \, \He_{12}(x) $ \\ \hline
			5 & $(\kappa_7/5040) \, \He_7(x) + (\kappa_6 \kappa_3/4320 + \kappa_5\kappa_4/2880)   \,  \He_9(x) + (\kappa_5 \kappa_3^2/8640 + \kappa_4^2\kappa_3/6912)   \,  \He_{11}(x) +  (\kappa_4 \kappa_3^2/31104)   \,  \He_{13}(x)  $ \\
			& $ + (\kappa_3^5/933120) \, \He_{15}(x) $
			\\ \hline
		\end{tabular}
		\caption{Edgeworth series terms}
		\label{tab:edgeworth}
	\end{table*}
	In general, the Edgeworth series terms can be written as \cite{blinnikov1998expansions}:
	\begin{equation}
		\begin{aligned}
			& p_E(x) = \frac{e^{-x^2/2}}{\sqrt{2\pi}} \bigg[ 1 + \\
			&\sum_{s=1}^\infty \sum_{\{k_m\} } \He_{s+2r}(x) \prod_{m=1}^s \frac{1}{k_m!} \left( \frac{\kappa_{m+2}}{(m+2)!} \right)^{k_m} \bigg],
		\end{aligned}
	\end{equation}
	where the summation over $\{k_m\}$ in the above denotes summatuion over all non-negative integer solutions of the Diophantine equation
	\begin{equation}
		k_1 + 2k_2 + \ldots + sk_s = s,
	\end{equation}
	and $r$ is the sum of these integers for each solution: $r = \sum k_m$. The explicit forms for a few of the orders of the Edgeworth expansion can be found in \cref{tab:edgeworth}

	\section{Kernel density approximation}\label{app:kde}
	Here, we give a quick overview of the kernel density approximation method. Supposing we have access to a finite number of samples drawn from a distribution function, the goal is to approximate the distribution function. To this end, a broadening kernel is placed at the position of each of the outcomes and a normalized sum approximates the underlying distribution:
	\begin{equation}
		\hat{p}(x) = \frac{1}{Mh} \sum_{i=1}^M K\left(\frac{x-X_i}{h}\right),
	\end{equation}
	where $K$ is a kernel (e.g. Gaussian, Lorentzian, etc.) with mean of 0 and variance of 1, $X_i, i=1,\ldots,M$ are the outcomes of sampling and $h$ is the broadening factor.

	The analysis of error in reconstructing the above QPE-kernel energy distribution with kernel density estimation follows a standard approach \cite{wand1994kernel}. First, the error is quantified by the quantity mean integrated square error (MISE):
	\begin{equation}
		\text{MISE} = \mathbb{E} \left( \int dx \; (\hat{p}(x) - p(x) )^2 \right),
	\end{equation}
	where $\hat{p}(x)$ is the approximated distribution for the underlying distribution $p(x)$. When a sample of size $M$ is used, it is well known that with an appropriate choice of $h$, i.e.~$h_{\text{opt}} \sim 1/M^{1/5}$, the error also shows the behavior $\text{MISE} \sim 1/M^{4/5}$.

	\section{Details of the quantum eigenvalue transformation of unitary matrices method}\label{app:QETU}

	\begin{figure}[t]
		\centering
		\includegraphics[width=0.5\textwidth]{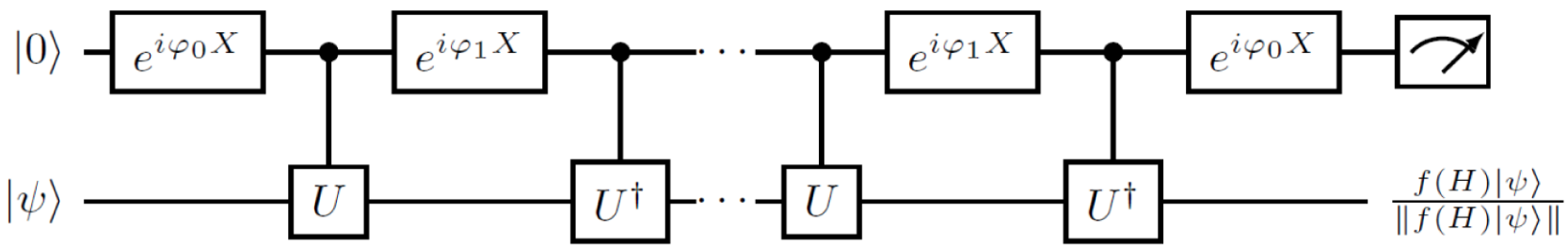}
		\caption{THE QETU circuit, see main text for explanation. The figure is taken from \cite{dong2022ground}.}
		\label{fig:qetu_circuit}
	\end{figure}

	The method consists of a quantum signal processing circuit \cite{low2017optimal,martyn2021grand} that implements a unitary matrix that block encodes a polynomial function $f(H) = P(\cos(H/2))$, where $H$ is the Hamiltonian of interest and $P$ is an even polynomial of degree $d$. A schematic of the quantum circuit is shown in Fig.~\ref{fig:qetu_circuit}. The circuit works by implementing $U=e^{-iH}$ and its Hermitian conjugate controlled on a single ancilla, a total number of $d$ times. The parameters $\varphi_0,\varphi_1,\ldots,\varphi_{d/2}$ are determined based on the polynomial of interest. Upon measuring the ancilla qubit at the end and obtaining the outcome $0$, the implementation has been successful, the probability of success is given by $\| P(\cos(H/2)) \ket{\psi} \|$.
	
	A scheme of the quantum circuit is shown in \cref{fig:qetu_circuit}. The circuit works by implementing $U=e^{-iH}$ and its Hermitian conjugate controlled on a single ancilla, a total number of $d$ times. The parameters $\varphi_0,\varphi_1,\ldots,\varphi_{d/2}$ are determined based on the polynomial of interest. Upon measuring the ancilla qubit at the end and obtaining the outcome $0$, the implementation has been successful, the probability of success is given by $\| P(\cos(H/2)) \ket{\psi} \|$.
	
\begin{figure}[!t]
		\centering
		\includegraphics[width=0.35\textwidth]{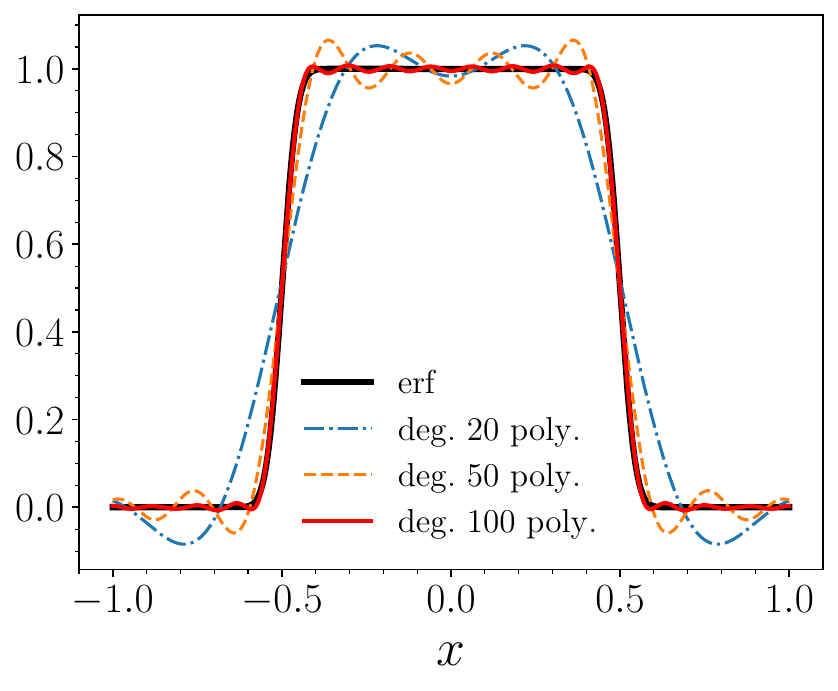}
		\caption{The even filtering functions required for QETU.  Polynomial approximations of the same function are also shown. }
		\label{fig:approx_poly_for_erf}
	\end{figure}

	The polynomial that needs to be implemented for our energy filtering task should be a symmetric function that retains low energies and filters high energies. 
	We take the spectrum of the Hamiltonian to lie within the interval $[-\pi+\eta,0-\eta]$, if necessary this can be done by adding a constant to and/or rescaling the Hamiltonian before performing QETU. Note that this is contrary to the original setting of Ref.~\cite{dong2022ground} (the spectrum is contained in $[\eta,\pi-\eta]$) and coarse QPE mentioned above; the reason for this change is better performance. 
	We need a polynomial $P$ which when expressed as $P(\cos(H/2))$ can filter high energies;
	it is straightforward to see that using the following combination of error functions, which we will try to imitate using the polynomial $P$, it is possible to filter high energies:
	\begin{equation}\label{eq:even_from_erf}
		\xi_{k,\mu}(x) = \frac12 \left[ \mathrm{erf}( - k (x-\mu) ) + \mathrm{erf}( k (x+\mu) ) \right],
	\end{equation}
	with $0<k$ and $0<\mu<1$ determining the steepness and position of the transitions, i.e.~position of energy filtering in the function. We use the prescription in Appendix A of Ref.~\cite{low2017hamiltonian} to reconstruct the error function $\mathrm{erf}(kx)$ in terms of Chebyshev polynomials as follows:
	\begin{equation}
		\begin{aligned}
			p_{\text{erf} , k , n} &= \frac{2k \, e^{-k^2/2}}{\sqrt{\pi}} \bigg( I_0(k^2/2) \\
			+ \sum_{j=1}^{(n-1)/2} & (-1)^j I_j(k^2/2)  \left[ \frac{T_{2j+1} (x)}{2j+1} - \frac{T_{2j-1} (x)}{2j-1} \right]\bigg),
		\end{aligned}
	\end{equation}
	where $T_j$ is the degree $j$ Chebyshev polynomial, $I_j$ is the modified Bessel function of the first kind. Note that $p_{\text{erf} , k , n}$ is an odd polynomial of degree $n$; it is the degree $n$ that controls the error in approximating $\mathrm{erf}(kx)$ and thus ensuring that low energies are retained and high energies filtered, and therefore it should be chosen large enough (see below). An example constructing polynomials like this is shown in \cref{fig:approx_poly_for_erf}.
	
	Applying a successful round of QETU filtering to a state $\ket{\psi} = \sum_E \varphi_E \ket{E}$, we end up with the following unnormalized state:
	\begin{equation}
		\sum_E \varphi_E \, P(\cos (E/2)) \ket{E}\ket{0},
	\end{equation}
	This shows that supposing we want to keep energies below $E_l$ and filter energies above $E_u$, we can choose a filtering function in \cref{eq:even_from_erf} (to be approximated by $P$) with $\mu = \frac{\cos(E_u/2)+\cos(E_l/2)}{2}$ and $\frac{1}{k} = \zeta \, \frac{\cos(E_u/2)-\cos(E_l/2)}{2}$. The factor $\zeta$ is added so that it is possible to control the intensity of filtering while keeping the degree of the polynomial and the cost down.

	The degree of a polynomial that needs to be used for this task will have a scaling $O(\Gamma^{-1} \log\epsilon^{-1})$, where $\epsilon$ is the error in the polynomial approximation and $\Gamma$ is the energy scale over which the transition in the error functions in \cref{eq:even_from_erf} happens, and thus should scale as $1/k$. 
	Apart form the above asymptotic scaling, in practice, we choose $n$ by examining how good of an approximation is achieved for degree $n$.

	\section{Simplified numerical example for the quantum refining step}\label{app:qetu_vs_cqpe}
	Here, as a simple concrete model, we consider a Gaussian energy distribution for our initial state. This Gaussian distribution can be characterized by a mean value $\bar{E}$ and a width $\sigma_E$:
	\begin{equation}\label{eq:gaussian_energy_distribution}
		A(E) = \frac1{\sqrt{2\pi} \sigma_E} \ e^{-\frac{(E-\bar{E})^2}{2\sigma_E^2} },
	\end{equation}
	Even though the energy distribution of an initial state might not actually be close to Gaussian in general, but we expect at least some variational states to show qualitatively similar behavior.

	We work with following concrete example of a Gaussian distribution: $\bar{E} = 0.06,\, \sigma = 0.02$.
	We would like to estimate the resources required to obtain a result close to 0 using QPE.
	With the above choice of parameters for the energy distribution, the accumulated weight below 0 is $p_<(0) = 0.0013$. This means that we need roughly $1/p_<(0)$ measurements to obtain a value around 0. We can perform quantum refining to decrease the number of times the most expensive quantum energy estimation routine is performed.
	We take this most precise routine to be a QPE with $k=10$ digits for this example, however, we tolerate an error of $2^{-8}$, discarding the last two digits in any QPE outcome. The number of digits $k$ furthermore determines the total evolution time required as $T \sim 2^{k}$. We characterize the cost of different operations by the number of queries they require to the unitary $e^{-iH}$, which for a $k$-digit QPE measurement becomes $2^k$. This means that each round of the ultimate QPE measurement brings in a cost of $2^{10}$.
	
	We first consider coarse QPE for energy filtering in this setting. 
	Concretely, we do a coarse QPE measurement with 4 digits and only keep the results that show outcome 0 in the measured phase register, we can filter out some part of the weight as shown in \cref{fig:filtering} (left).
	This outcome happens with a probability of $W_{k=4} = 0.10$. In this new energy distribution the total weight below zero now reads $p'_{\text{QPE},<}(0) = 0.012$.
	This means that, after such measurement, roughly ten times less rounds of the precision QPE will be required compared to the initial state. 
	
	One can do this procedure one more time with a coarse QPE with 5 digits now and postselecting on the outcome 0 again; the resulting weight distribution can be seen in \cref{fig:filtering} (left). The probability of such outcome (given the previous outcome of 0 with 4 digits) is now $W_{k=4;5} = 0.13$ (this means that the probability of obtaining 0 in the 4-digit measurement and then also 0 in the 5-digits measurement is $W_{k=4}W_{k=4;5} = 0.013$). Remarkably with this measurement, the total weight below zero becomes $p''_{\text{QPE},<}(0) = 0.083$; this results in close to two orders of magnitude decrease in the number of precision QPE measurement required for obtaining outcomes close to 0. This is achieved for a cost of $2^4+2^5$ which is an insignificant overhead compared with the cost of the most precise QPE measurement.

	For QETU, we shift the energies so that low energies are located close to $-\pi$ as discussed in \cref{app:QETU}.
	We use a degree 200 polynomial to approximate a step function as shown in \cref{fig:filtering}. The normalized distribution after QETU has been performed moves to the left and thus some of the higher energies are filtered.
	Probability of success in this case is $W_{\text{QETU}} = 0.21$ and the total weight below 0 after the procedure can also be calculated as $p'_{\text{QETU},<} = 0.0056$. This means decreasing the number of repetitions roughly by a factor of 4. As the polynomial that is used is order 200, the number of required queries to $e^{-iH}$ is also 200. 
	
	We see that both coarse QPE and QETU refining methods can be helpful for a cost that is an insignificant fraction of the ultimate QPE cost, but coarse QPE acts considerably better for a lower cost. As creating steep polynomials like the one used here is generally a hard task, we believe this result should hold generically even though we tested it here for a simple model.

	\subsubsection{Mitigating the leakage}
	Another thing which can be studied in this simple model is the probability of leakage before and after the refining is performed. We only consider coarse QPE for this. 
	Before any of the measurements are performed the total probability of leakage is equal to $p_{\text{leak}} = 0.00097$ which close to $p_<(0) = 0.0013$ and this can be problematic by contributing outcomes below the actual energy levels of th system. Upon performing 4-digit and 5-digit QPE measurements discussed above, the probability of leakage become $p'_{\text{leak}} = 0.0019$ and respectively $p''_{\text{leak}} = 0.0036$. These two values when compared with $p'_{\text{QPE},<}(0) = 0.012$ and $p''_{\text{QPE},<}(0) = 0.083$ show that the probability of leakage has decreased substantially enough compared to the probability of obtaining results of interest, so that its occurrence has become improbable, and thus quantum refining has suppressed the possibility of leakage also.

	\section{Error analysis of the QPE leakage approximate form}\label{app:leakage}
	In this appendix we analyze \cref{eq:total_probability_leakage_sum} and in particular how the approximation in \cref{eq:single_level_approx} can be performed. 
	\begin{equation}
		p_{\text{leak}}(E_n) =  \sum_{x_j<x_{\text{upper}}} \frac{1}{2^{2k}} \left(\frac{\sin^2 (\pi \delta_n)}{ \sin^2 \left(\frac{\pi}{2^k} [x_n+\delta_n - x_j]\right)} \right),
	\end{equation}
	we take the lower bound in the summation over $x_j$ to be $-2^{k-1}$, and we are using the periodicity of the QPE results.
	A lower bound and an upper bound for the above sum can be found by using the following integral form:
	\begin{equation}
		\begin{aligned}
			I(x_0) &= \frac{\sin^2(\pi \delta_n)}{2^{2k}} \, \int_{-2^{k-1}}^{x_0}  \frac{dx}{\sin^2(\frac{\pi}{2^k} (x_n+\delta_n-x))} \\
			&= \frac{\sin^2(\pi \delta_n)}{2^{2k}} \left[ \left. \frac{2^k}{\pi} \cot \left(\frac{\pi}{2^k} (x_n+\delta_n-x) \right)\right|^{x_0}_{-2^{k-1}} \right] .
		\end{aligned}
	\end{equation}
	It is easy to check that:
	\begin{equation}
		I(x_{\text{upper}}-1) \leq p_{\text{leak}}(E_n) \leq I(x_{\text{upper}}).
	\end{equation}
	This readily results in \cref{eq:single_level_approx}, and the error can also be shown to have the form $O\left(\max \left[2^{-2k},(x_n-x_{\text{upper}})^{-2} \right] \right)$  by evaluating $I(x_{\text{upper}}) - I(x_{\text{upper}}-1)$.
	
	\bibliography{initial.bib}
	
\end{document}